\documentclass[pra,twocolumn,longbibliography,preprintnumbers,superscriptaddress,amsmath,amssymb,floatfix]{revtex4-2}
\usepackage{graphicx}
\usepackage{subfigure}
\usepackage{mathrsfs}
\usepackage{amsfonts}
\usepackage{times}
\usepackage{amsmath}
\usepackage{amsthm}
\usepackage{leftidx}
\usepackage{xcolor}
\DeclareMathOperator*{\argmin}{argmin}
\DeclareMathOperator*{\argmax}{argmax}
\usepackage{chngcntr}
\usepackage{color}
\usepackage[colorlinks,linkcolor=blue,citecolor=blue]{hyperref}

\usepackage{bbold}
\usepackage{braket}
\usepackage{mathtools}
\usepackage{siunitx}
\usepackage{CJKutf8}

\begin{document}
\title{Maxwell's demon for quantum transport}
\author{Kangqiao Liu}
\email{kqliu@xhu.edu.cn}
\affiliation{School of Science, Key Laboratory of High Performance Scientific Computation, Xihua University, Chengdu 610039, China}
\affiliation{Department of Physics, The University of Tokyo, 7-3-1 Hongo, Bunkyo-ku, Tokyo 113-0033, Japan}
\author{Masaya Nakagawa}
\affiliation{Department of Physics, The University of Tokyo, 7-3-1 Hongo, Bunkyo-ku, Tokyo 113-0033, Japan}
\author{Masahito Ueda}
\affiliation{Department of Physics, The University of Tokyo, 7-3-1 Hongo, Bunkyo-ku, Tokyo 113-0033, Japan}
\affiliation{RIKEN Center for Emergent Matter Science, 2-1, Hirosawa, Wako-shi, Saitama, 351-0198, Japan}
\affiliation{Institute for Physics of Intelligence, The University of Tokyo, 7-3-1 Hongo, Bunkyo-ku, Tokyo 113-0033, Japan}
\date{\today}

\begin{abstract}
While most of the existing quantum information engines assisted by Maxwell's demon harness thermal fluctuations, those that rectify only quantum fluctuations have recently been constructed. We propose an alternative type of quantum information engine that harnesses only quantum fluctuations to achieve cumulative energy storage and unidirectional transport of a particle. This unidirectional transport makes a stark contrast with the case without Maxwell's demon where the motion of a particle is confined to a finite region due to Bloch oscillations. We find a tradeoff relationship between the maximum power and the maximum velocity. With an improved definition of efficiency that includes all possible energy flows in the engine cycle, we numerically demonstrate the absence of a tradeoff relationship among power, efficiency, and power fluctuations, that is present for classical heat engines and classical information engines. We also evaluate the influence of experimentally unavoidable measurement imprecision on the performance of the quantum Maxwell's demon.
\end{abstract}
\maketitle

\section{Introduction}\label{sec: intro} 
The extractable work of a classical heat engine (CHE) is fundamentally limited by the second law of thermodynamics. The Szilard engine is the first classical information heat engine (CIHE) that utilizes Maxwell's demon, who can perform measurement and feedback (FB) control, to seemingly extract more work by harnessing thermal fluctuations than the second-law constraint, where the extra work gain is compensated for by the energy cost of information processing \cite{Szilard1964decrease, Brillouin1951maxwell, Landauer1961irreversibility, Bennett1982thermodynamics, Sagawa2008second, Sagawa2009minimal, Sagawa2011erratum}. Quantum information heat engines (QIHEs) have been proposed theoretically \cite{Kim2011quantum,Park2013heat,Brandner2015coherence,Chapman2015how,Yi2017single,Morikuni2017quantum,Ding2018measurement,Bozkurt2018work,Buffoni2019quantum,Seah2020maxwell,Poulsen2022quantum,Bhandari2023quantum,Bozkurt2023topological} and realized experimentally \cite{Camati2016experimental,Cottet2017observing,Masuyama2017information,Naghiloo2018information,Wang2018realization,Najera-Santos2020autonomous,Hern2022autonomous,Yan2022verification,Yu2024Experimental}. However, most of them extract work by rectifying thermal fluctuations and the role of quantum fluctuations remains elusive.

Quantum fluctuations, by which we mean the intrinsic randomness of measurement outcomes for quantum states, are unique to quantum systems. Recently, a minimal model of quantum information engines that rectifies only quantum fluctuations has been proposed \cite{Alexia2017engine}, where the working agent gains energy directly from quantum measurements on observables that do not commute with the system Hamiltonian. We call this type of quantum information engine a genuinely quantum information engine (GQIE) because thermal effects are absent.

If one can also store the extracted energy to transport a particle uphill against a potential barrier, then the stored energy can be utilized on demand
\cite{Toyabe2010,Admon2018experimental,Sahae2023356118}. From an energy-storage viewpoint, the energy increase can be regarded as charging a quantum battery \cite{Yan2023charging,Campaioli2024colloquium}. This non-cyclic transport engine is crucially important, for example, in molecular motors \cite{Leigh2003unidirectional,Astumian2007design,Greb2014light,Erbas2017rotary,Kassem2017artificial,Feng2021molecular} aside from the interest in transport phenomena themselves \cite{Hanggi2009artificial,Laird2015quantum}. A prototypical example of a transport CIHE was realized experimentally with a Brownian particle by Toyabe \textit{et al.} \cite{Toyabe2010} and its power, transport velocity, and efficiency have been investigated in Refs.~\cite{Admon2018experimental,Kiran2022driven,Sahae2023356118}. Recently, a transport GQIE has been proposed \cite{Elouard2018efficient}. However, a heat bath is used to prepare an equilibrium state in each cycle with unknown thermalization times, which makes it difficult to evaluate its power and velocity.

\begin{figure}[t!]
\begin{center}
\includegraphics[width=\columnwidth]{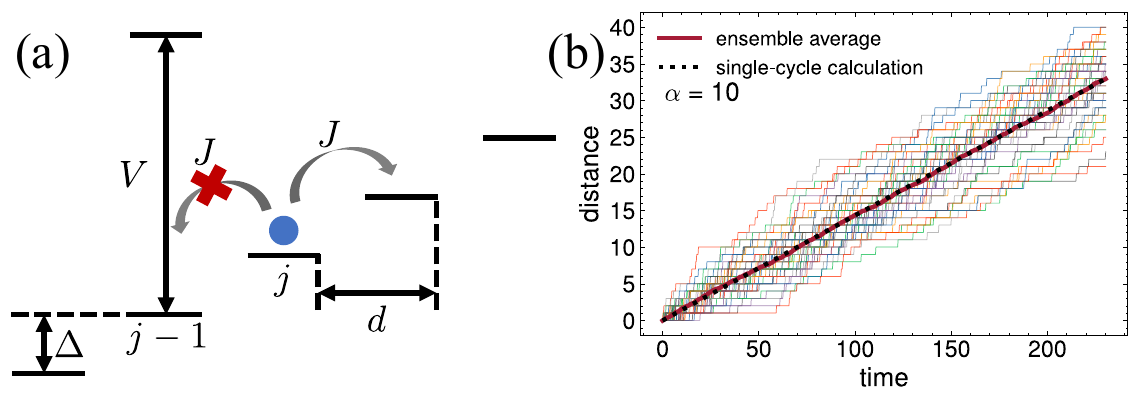}
\end{center}
\caption{(a) Setup. A particle hops on a one-dimensional tilted lattice with lattice constant $d$, hopping amplitude $J$, and step height $\Delta$. If a projective position measurement finds the particle at site $j$, the potential at site $j-1$ is instantly raised by $V\gg \Delta, J$ to prevent the particle from hopping downwards. The particle then evolves according to Eq.~\eqref{eq: our limiting H} during time $t$. Then we perform the projective position measurement again which may find a particle at a higher position $j'>j$. Through repetitive measurement and feedback, the particle climbs up the stairs to gain potential energy. (b) Numerical simulations of the traveled distance for 100 realizations (thin colored), each of which runs 1000 engine cycles with $\alpha:=\Delta/J =10$ and $\tilde{V}=10^3$. The single-cycle optimization (black dotted) agrees well with the trajectory ensemble average (thick red).}
\label{fig: setup}
\end{figure}

In this paper, we propose a quantum counterpart of Toyabe's CIHE, i.e., a GQIE with a tilted one-dimensional (1D) lattice that can store energy cumulatively without resorting to thermalization and allows us to introduce well-defined power and velocity. The achieved unidirectional transport is highly nontrivial because, without Maxwell's demon, a particle in this lattice undergoes Bloch oscillations which prohibit transport \cite{Bloch1929oscillation,Hartmann2004dynamics}. We find that the maximum power and velocity obey a tradeoff relationship, implying that a larger power can be obtained at the expense of a smaller velocity and vice versa. We define its efficiency by considering both the energy costs of measurement and information erasure \cite{Sagawa2009minimal,Sagawa2011erratum}. We report the absence of a tradeoff relationship among power, efficiency, and power fluctuations, which is present for CHEs \cite{Pietzonka2018universal} and CIHEs \cite{tanogami2023universal}. We study the influence of measurement imprecision \cite{wiseman_milburn_2009book}, which is relevant to experiments \cite{Gross2021quantum}.

\section{Setup}\label{sec: setup} 
A particle hops on a tilted 1D lattice as schematically illustrated in Fig.~\ref{fig: setup}(a). It is described by the Wannier-Stark (WS) Hamiltonian \cite{GLUCK2002Wannier,Hartmann2004dynamics,ashcroft2022solid}, 
\begin{equation}\label{eq: WS Hamiltonian}
H_{\rm WS} = \sum_{n=-\infty}^{+\infty}-J(\ket{n}\bra{n+1}+\ket{n+1}\bra{n}) + \Delta n\ket{n}\bra{n}, 
\end{equation}
where $J$ is the hopping amplitude, $\Delta:=Fd>0$ is the step height, $F$ is the gradient force, $d$ is the lattice constant, and $\{\ket{n}\}$ are the localized Wannier states \cite{Kohn1959analytic,Wannier1960wave,Nenciu1983existence}. This Hamiltonian has been widely implemented and utilized in experiments of cold atoms in optical lattices \cite{Simon2011quantum,Fukuhara2013quantum,Preiss2015strongly,Scherg2021observing,Natale2022Bloch}. It is an approximation to a continuous-space Hamiltonian $H_{\rm WS} = -\hbar^2\partial_x^2 /2m + U(x) + Fx$, where $U(x+d) = U(x)$ is a 1D periodic potential. We aim to pump up the particle against the linear potential to accumulate energy in its potential energy and achieve directed motion through the following four steps.

\noindent \textit{Step 1 -- measurement.} The initial state is a pure state at the first cycle or the time-evolved state of the previous cycle for later cycles. Projective position measurement is performed with the measurement operator $M_j = \Pi_j :=  \ket{j}\bra{j}$.

\noindent \textit{Step 2 -- feedback control.} If the particle is detected at site $j$, the potential at site $j-1$ is instantly raised by $V\gg \Delta, J$ to prevent the particle from hopping downwards, resulting in a new Hamiltonian $H_j := H_{\rm WS} + V\Pi_{j-1}$. The FB is ideally cost-free because the external work performed on the particle by this instantaneous feedback quench is
\begin{equation}
    W_{\mathrm{FB}}(j) := \bra{j}(H_j - H_{\mathrm{WS}})\ket{j},
\end{equation}
which vanishes for an ideal projective position measurement \cite{Sagawa2008second, Sagawa2009minimal, Sagawa2011erratum, Parrondo2015thermodynamics, Elouard2018efficient, Alexia2021short, Bhandari2023quantum}.

\noindent \textit{Step 3 -- unitary evolution.} The particle evolves unitarily during time $t$, which is a free parameter, with the new Hamiltonian $H_j$. A sufficiently large $V$ effectively imposes a rigid wall at site $j-1$ such that the wave function vanishes $\forall n\le j-1$, giving an effective Hamiltonian
\begin{equation}\label{eq: our limiting H}
    H_j=-J\sum_{n=j}^{N-1}\left(\ket{n}\bra{n+1}+\ket{n+1}\bra{n}\right)+\Delta\sum_{n=j}^{N}n\Pi_{n},
\end{equation}
where the chain is truncated at site $N$ for numerical simulations. The truncation is valid if $N \gg \langle n \rangle$, where $ \langle n \rangle$ is the distance traveled to be defined later. This semi-infinite chain description is justified by the Schrieffer-Wolff transformation in the large-$V$ limit and the requirement on the largeness of $V$ is also given in Appendix~\ref{app:large V truncation}.

\noindent \textit{Step 4 -- information erasure.} We erase the information of measurement outcomes stored in the probe's memory and start the next cycle.

\section{Power and velocity}\label{sec: p and v def} 
We are interested in the average energy storage rate (power $p$), the particle transport rate (velocity $v$), and the efficiency $\eta$ which is defined later. 
Without loss of generality, the particle is initially localized at site $0$ with a rigid wall placed at site $-1$, governed by the Hamiltonian $H_0$. The average distance traveled by the particle per cycle in units of $d$ is defined as $\langle n(t) \rangle := \bra{\psi(t)}\sum_{n=0}^{N} n\Pi_n \ket{\psi(t)} = \sum_{n=0}^N n P(n,t)$, where $P(n,t) := |\braket{n|\psi(t)}|^2$ is the probability of the particle being found at site $n$ at time $t$ and $\ket{\psi(t)} := \exp(-i H_0 t/\hbar)\ket{0}$ is the time-evolved state. The average energy gain per cycle, power, and transport velocity are defined as $E(t) := \Delta \langle n(t) \rangle$, $p(t) := E(t)/t$, and $v(t) := \langle n(t) \rangle/t = p(t)/\Delta$, respectively.

\begin{figure}[t!]
\begin{center}
\includegraphics[width=\columnwidth]{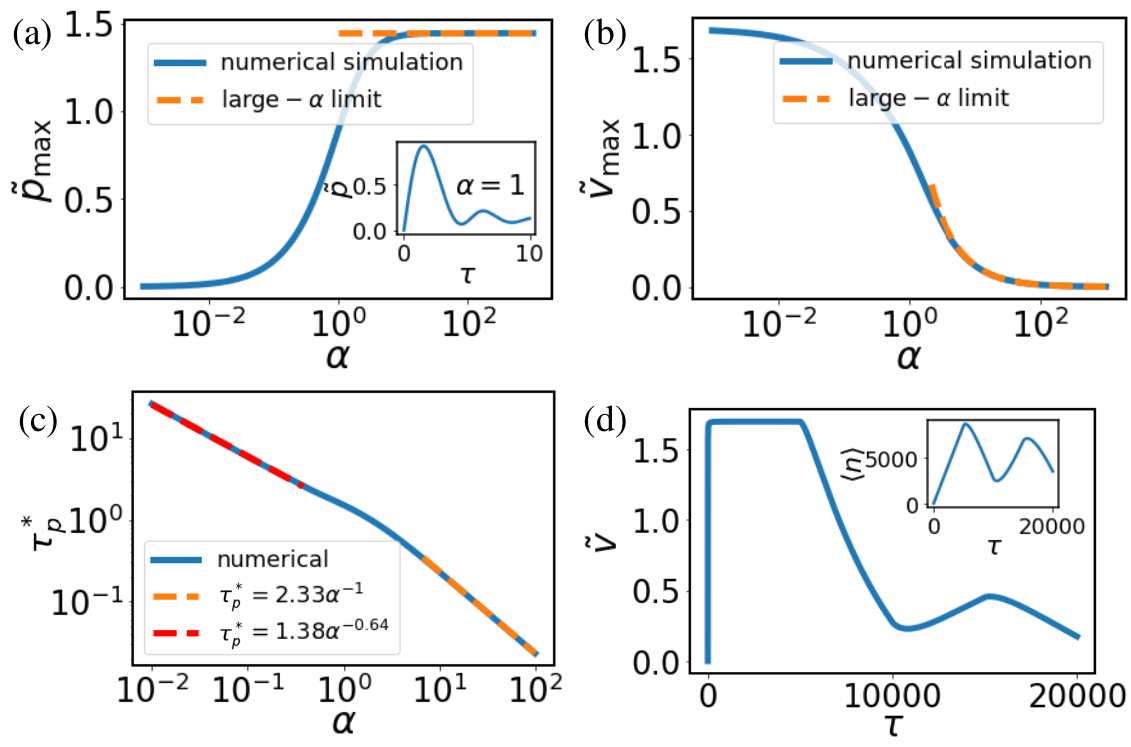}
\end{center}
\caption{(a) Maximal power $\tilde{p}_{\rm max}(\alpha)$ and (b) maximal velocity $\tilde{v}_{\rm max}(\alpha)$ as functions of gradient $\alpha$. The blue curves are numerical results and the orange dashed curves show the asymptotic results for large $\alpha$ (see Sec.~\ref{sec: p and v def}). The inset in (a) shows $\tilde{p}(1,\tau)$ as a function of $\tau$, indicating a well-defined optimal time. (c) Optimal time $\tau^*_p(\alpha)$ for $\tilde{p}_{\rm max}$ and $\tilde{v}_{\rm max}$. The orange-dashed curve is predicted by the large-$\alpha$ theory (see Sec.~\ref{sec: p and v def}). Data fitting (red) gives $\tau^* \propto\alpha^{-0.64}$ for small $\alpha$. (d) Velocity for $\alpha=0$ with $N=10000$ sites. The inset shows the displacement $\langle n(\tau) \rangle$. Both insets are plotted in linear scales. After a negligible transient stage, the velocity reaches about 1.7 until the wave packet reaches the right boundary as plotted in the inset. }
\label{fig: maximum p and v}
\end{figure}

We argue that these quantities can be studied within a single cycle because all cycles undergo the same dynamics except for a shift in the site index. For a trajectory consisting of $M$ FB cycles, the total accumulated energy gain is given by $W:=\sum_{i=1}^M w_i$, where a stochastic variable $w_i$ denotes the energy stored in the $i$th cycle. This stochastic energy gain is defined as follows. Suppose that at the $(i-1)$th cycle the outcome of the projective measurement is $n$, so that the particle's potential energy is $n\Delta$; in the next $i$th cycle, the measurement gives an outcome $n+l$ ($l\in\mathbb{N}$) and the potential energy becomes $(n+l)\Delta$, thus giving the energy gain of $w_i = l\Delta$. The values of $w_i$ and $w_j$ for $i\ne j$ are independent because the rank-1 measurement eliminates the quantum coherence in the measured state and the time evolution is the same except for a shift in the site index. For any cycle, $w_i$ has the same average value $E$ defined previously. In the limit of $M\to+\infty$, the power of a long trajectory converges to $p=\lim_{M\to+\infty}\sum_{i=1}^{M}w_i/Mt=M E/Mt=E/t$ due to the law of large numbers, i.e.,  $\lim_{M\to+\infty}\sum_{i=1}^{M}w_i=\lim_{M\to+\infty}M E$. Therefore, it is sufficient to study the performance within one cycle.

We define the dimensionless gradient $\alpha:= \Delta/J$, the evolution time $\tau:= Jt/\hbar$, and the dimensionless Hamiltonian $\tilde{H}_0:=H_0/J$. The performance of the engine is measured by $\tilde{E}(\alpha, \tau) := E(t)/J$ for energy, $\tilde{p}(\alpha, \tau):= \tilde{E}(\alpha, \tau)/\tau = p\hbar/J^2$ for power, and $\tilde{v}(\alpha, \tau) := \langle n(t) \rangle/\tau = v\hbar/J = \tilde{p}(\alpha, \tau)/\alpha$ for velocity. We study the dynamics numerically with two tunable parameters $\alpha$ and $\tau$. We take $N = 1000$ so that the effect of the right boundary can be neglected, except for Fig.~\ref{fig: maximum p and v}(d) where $N=10000$.

The maximal power for a fixed value of $\alpha$ is defined as
\begin{equation}\label{eq: def of p_max}
    \tilde{p}_{\rm max}(\alpha) := \max_{\tau} \tilde{p}(\alpha,\tau) = \tilde{p}\left(\alpha,\tau^*_p(\alpha)\right),
\end{equation}
where the optimal time is $\tau^*_p(\alpha):= \argmax_{\tau} \tilde{p}(\alpha,\tau)$; see the inset of Fig.~\ref{fig: maximum p and v}(a) as an example. Because $\tilde{p} =\alpha \tilde{v}$, we have $\tau^*_v(\alpha) = \tau^*_p(\alpha)$. The maximum velocity is
\begin{equation}\label{eq: def of v_max}
    \tilde{v}_{\rm max}(\alpha) := \max_{\tau} \tilde{v}(\alpha,\tau) = \tilde{v}\left(\alpha,\tau^*_p(\alpha)\right) = \frac{\tilde{p}_{\rm max}(\alpha)}{\alpha}.
\end{equation}

Figures~\ref{fig: maximum p and v}(a) and (b) plot $\tilde{p}_{\rm max}(\alpha)$ and $\tilde{v}_{\rm max}(\alpha)$, respectively. We see that $\tilde{p}_{\rm max}$ increases monotonically, as opposed  to $\tilde{v}_{\rm max}$, indicating a tradeoff that a large power is attained at the expense of a suppressed velocity and vice versa. This is our first main observation. In Fig.~\ref{fig: maximum p and v}(c), the optimal time $\tau^*_p(\alpha)$ is plotted as a function of $\alpha$. The particle is effectively described by a two-site system for large $\alpha$, and for small $\alpha$ the particle travels almost on a flat chain with an initial velocity 1.7 due to the Heisenberg’s uncertainty relation associated with the initial localization as shown in Fig.~\ref{fig: maximum p and v}(d). We provide detailed analyses below.

In the large-$\alpha$ regime, the dynamics is governed by an effective Hamiltonian $\tilde{H}_{\rm eff} = -(\ket{0}\bra{1}+ \ket{1}\bra{0}) + \alpha \ket{1}\bra{1}$. At time $\tau$, the state $\ket{\psi(\alpha,\tau)} =\exp(-i\tilde{H}_{\mathrm{eff}}\tau)\ket{0}$ can be given analytically as $\ket{\psi(\alpha,\tau)} = c_0 (\alpha,\tau) \ket{0} + c_1 (\alpha,\tau) \ket{1}$, where
\begin{align}
    & c_0 (\alpha,\tau) = \frac{e^{-i\frac{\alpha \tau}{2}}}{\sqrt{4+\alpha^2}}\left[ \sqrt{4+\alpha^2}\cos\left(\frac{\sqrt{4+\alpha^2}\tau}{2}\right) \right. \nonumber\\
    & \qquad \qquad\ + \left. i \alpha \sin \left(\frac{\sqrt{4+\alpha^2}\tau}{2}\right) \right], \\
    & c_1 (\alpha,\tau) = \frac{2i e^{-i\frac{\alpha \tau}{2}}}{\sqrt{4+\alpha^2}}\sin \left(\frac{\sqrt{4+\alpha^2}\tau}{2}\right).
\end{align}
Therefore, the expected value of the energy gain is $\tilde{E}(\alpha,\tau) = \alpha |c_1 (\alpha,\tau)|^2 = 4\alpha\sin^2 (\sqrt{4+\alpha^2}\tau/2)/(4+\alpha^2)$. Because $\alpha \gg 1$, it can be simplified as $\tilde{E}(\alpha,\tau) \approx 4\sin^2 (\alpha \tau/2) /\alpha$. Dividing it by $\tau$ yields the power as $\tilde{p}(\alpha,\tau) = 4\sin^2 (\alpha \tau/2)/\alpha\tau$ which attains its maximal value $\tilde{p}_{\rm max}(\alpha) \approx 1.44$ at $\tau^*_p(\alpha) \approx 2.33/\alpha$. The maximum velocity is given by $\tilde{v}_{\rm max}(\alpha) = \tilde{p}_{\rm max}(\alpha)/\alpha \approx 1.44/\alpha$ which is obtained for the same operation time, in excellent agreement with the numerical results in Fig.~\ref{fig: maximum p and v}. The reason is that although the probability of the particle hopping to the right neighboring site is small, once it hops, a large amount of energy $\alpha$ is gained in the potential energy.

In the small-$\alpha$ regime, we provide a qualitative explanation to the saturation of $\tilde{v}_{\rm max}$ at a value of about 1.7 in Fig.~\ref{fig: maximum p and v}(b). We consider a flat lattice in the $\alpha\to 0$ limit. Initially, the particle is localized at site $0$. After a short transient dynamics about $\tau =  O(1)$, the particle travels in the positive direction at a constant velocity as shown in Fig.~\ref{fig: maximum p and v}(d). For a long but finite chain as used in the simulation, the wave function will eventually be reflected back by the right boundary. We use an $N=10000$ chain and let the dynamics run for $\tau = 20000$. As shown in the inset of Fig.~\ref{fig: maximum p and v}(d), the traveled distance $\langle n(\tau) \rangle$ first linearly increases with $\tau$ except for a negligible transient period, and then decreases due to the reflection when it reaches $\langle n(\tau) \rangle =10000$. The velocity is kept constant before the wave function reaches the right boundary. The physics underlying the value of this constant velocity is attributed to Heisenberg's uncertainty principle. Since the particle is initially localized, its momentum undergoes significant quantum-mechanical fluctuations which provide the initial energy for the particle to travel to the right. Because the initial kinetic energy $K$ has a zero mean, its standard deviation is given by $\sigma(K)=\sqrt{\bra{\psi(0)}H_0^2\ket{\psi(0)}} = J$. Therefore, the velocity is expected to be about $\tilde{v}=v\hbar/Jd=(\hbar/Jd)\sqrt{2 \sigma(K)/m^*} = 2$ with $m^*= \hbar^2/2Jd^2$ being an effective mass of the particle \cite{ashcroft2022solid}. This value is close to the numerically obtained result of 1.7.

\begin{figure*}[t!]
\begin{center}
\includegraphics[width=\textwidth]{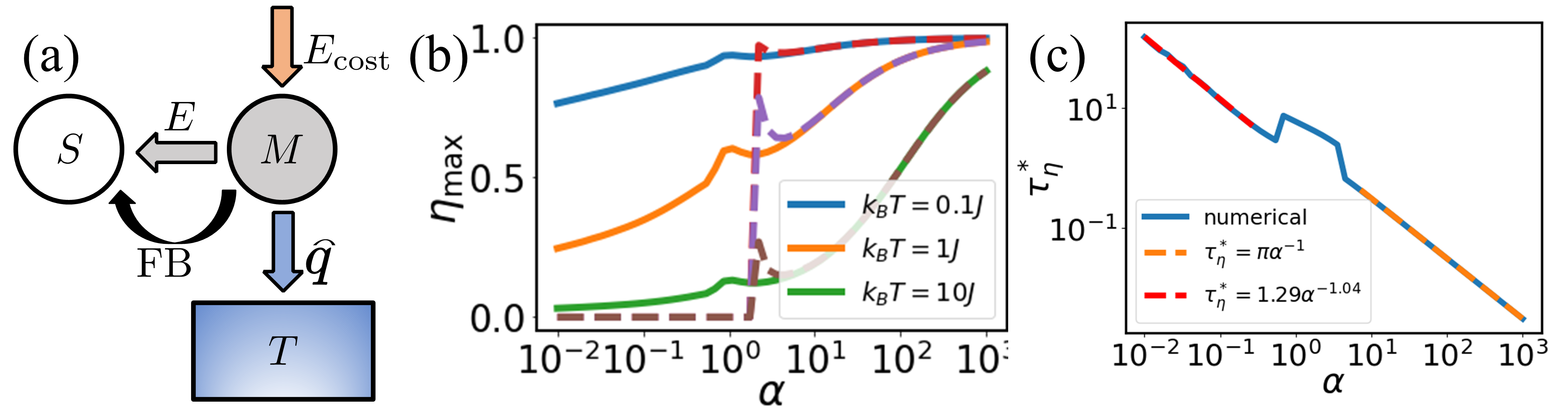}
\end{center}
\caption{(a) Energy flow diagram for our engine. The probe $M$ plays the role of an agent because it is cyclic. The total energy input $E_{\rm cost}$ consists of the costs of measurement and information erasure, whose resource is not a heat bath. Heat $q:= E_{\rm cost} - E$ is dissipated into a heat bath at temperature $T$ to erase the information in $M$. The FB stores energy $E$ in the potential energy of the particle $S$. (b) The maximal efficiency $\eta_{\rm max}(\alpha)$ for $\tilde{T} = 0.1$ (blue), $1$ (orange), and $10$ (green). The dashed curves show the large-$\alpha$ approximation. (c) Optimal time $\tau^*_{\eta}(\alpha)$ with $\tilde{T}=1$. The orange dashed line is the large-$\alpha$ calculation and data fitting (red) for the small-$\alpha$ region gives an almost inverse linear relation.}
\label{fig: efficiency}
\end{figure*}

\section{Efficiency}\label{sec: efficiency}
The efficiency of a CHE is defined as the ratio of the output work to the heat absorbed from a hot heat bath. We define analogously the efficiency $\eta$ of our engine by clarifying the energy flows in Fig.~\ref{fig: efficiency}(a). A crucial difference from a CHE is the origin of the total energy input which involves two contributions. (1) In Step 1, the measurement on average increases the internal energy of the particle [$S$ in Fig.~\ref{fig: efficiency}(a)] by $E$, requiring an energy input $E_{\rm meas}$. (2) In Step 4, the state of the probe [$M$ in Fig.~\ref{fig: efficiency}(a)] is reset to its default value by discarding the information about the measurement outcomes stored in its memory through coupling with a heat bath at temperature $T$ in Fig.~\ref{fig: efficiency}(a) according to Landauer's protocol \cite{Piechocinska2000information,Rio2011thermodynamic,Berut2012experimental}. This requires an energy input $E_{\rm eras}$. The total energy input is $E_{\rm cost} = E_{\rm meas} + E_{\rm eras}$. We note that the energy resource is not a heat bath. 

It is proved that a fundamental minimal total energy cost $E_{\rm cost}$ exists, whereas no fundamental lower bound exists for each individual process of measurement and erasure \cite{Sagawa2009minimal,Sagawa2011erratum}. This implies the necessity of considering both energy inputs to define efficiency. In Refs.~\cite{Alexia2017engine, Elouard2018efficient, Alexia2021short}, $E_{\rm meas}$ is not considered as an input, while in Refs.~\cite{Brandner2015coherence,Yi2017single}, $E_{\rm eras}$ is not included. In the former, the efficiency is defined as $\eta' := (E-E_{\rm eras})/E$. However, information erasure can, in principle, be performed without work \cite{Sagawa2009minimal,Vaccaro2011information}, resulting in $\eta' = 1$ regardless of the details of the engine. On the other hand, a simple replacement of $E_{\rm eras}$ by $E_{\rm cost}$ does not give a proper definition because for all cases $\eta'' := (E-E_{\rm cost})/E <0$.

We therefore define the efficiency of our engine as
\begin{equation}\label{eq: def of eta}
    \eta(\Delta, t) := \frac{E}{E_{\rm cost}} = \frac{E(\Delta, t)}{E(\Delta, t) + k_B T H(\Delta, t)},
\end{equation}
where we use standard techniques in information thermodynamics \cite{Jacobs2009second, Sagawa2009minimal, Sagawa2011erratum, Jacobs2012quantum,Faist2015minimal, Abdelkhalek2016quantum, Deffner2016quantum, Mohammady2021classicality} to get $E_{\rm cost} = E + k_B T H$ \cite{Sagawa2009minimal, Sagawa2011erratum,Abdelkhalek2016quantum} with $H(\Delta, t):= -\sum_{n} P(n) \ln P(n)\ge 0$ being the Shannon entropy of the probability distribution $\{P(n)\}$ of measurement outcomes. The definition \eqref{eq: def of eta} ensures $\eta\in [0,1]$. During the erasure process, heat $q:= E_{\rm cost} - E = k_B T H$ is dissipated into the heat bath. We have $\eta(\alpha, \tau) = \tilde{E}(\alpha, \tau)/[\tilde{E}(\alpha, \tau) + \tilde{T} H(\alpha, \tau)]$, where $\tilde{T} := k_B T /J$. With Eq.~\eqref{eq: def of eta}, the value of the optimal efficiency in Refs.~\cite{Alexia2017engine,Elouard2018efficient,Alexia2021short} and the corresponding conditions remain unchanged while avoiding unphysical consequences discussed in the last paragraph.

The solid curves in Fig.~\ref{fig: efficiency}(b) show the maximal efficiency
\begin{equation}\label{eq: def of eta max}
    \eta_{\rm max}(\alpha) := \max_{\tau} \eta(\alpha, \tau) = \eta\left(\alpha, \tau^*_{\eta}(\alpha)\right)
\end{equation}
for dimensionless temperature $\tilde{T}=0.1$, 1, and 10. The optimal time $\tau^*_{\eta}(\alpha)$ is plotted in Fig.~\ref{fig: efficiency}(c). The maximal efficiency increases almost monotonically and saturates at unity with a small bump in the middle. The bump is attributed to the complicated quantum interference between Bloch oscillations and a wave reflected by the left rigid wall, which coherently shrinks the spread of the wave packet with some irregularity (see Appendix~\ref{app: bump} for details).
In the large-$\alpha$ regime, we have  $\eta(\alpha,\tau) = 4\sin^2(\alpha\tau/2) /[ 4\sin^2(\alpha\tau/2) + \alpha\tilde{T} H(4\sin^2(\alpha\tau/2)/\alpha^2) ]$. The maximum $\eta$ is obtained at $\tau^*_{\eta}(\alpha)=(2k-1)\pi/\alpha$ where $k\in \mathbb{N}$, giving
\begin{equation}
    \eta_{\rm max}(\alpha) = \frac{1}{1+\frac{\tilde{T}}{4}\alpha H\left(\frac{4}{\alpha^2}\right)},
\end{equation}
which approaches unity because $\alpha H(4/\alpha^2) \to 8\ln(\alpha)/\alpha\to 0$. It is plotted in dashed curves in Fig.~\ref{fig: efficiency}(b) and agrees well with the numerical results where a spurious divergence occurs at $\alpha=2$, indicating the breakdown of the large-$\alpha$ approximation. This unit efficiency is similar to that in Refs.~\cite{Alexia2017engine,Elouard2018efficient} because $H$ is so small that the dissipated heat $q$ is small. The large-$\alpha$ approximation of $\tau^*_{\eta}$ also agrees with Fig.~\ref{fig: efficiency} (c). The unit efficiency is our second main observation.

\begin{figure*}[t!]
    \centering
    \includegraphics[width=\textwidth]{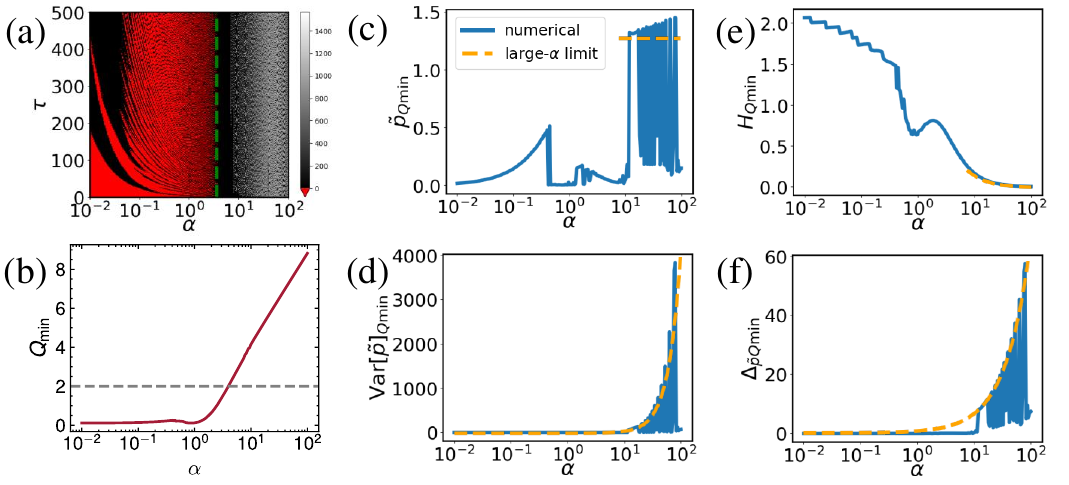}
    \caption{(a) Numerical calculation of $Q(\alpha,\tau)-2$ defined in inequality~\eqref{eq: TUR to test} for $\alpha\in[10^{-2},10^2]$ and $\tau\in [0,500]$. Data points representing $Q-2<0$ are plotted in red and nonnegative points are shown in gray, where whiter colors represent larger values. A clear boundary near $\alpha=3.6$ that separates the two regions is indicated by the dashed line; to the left of it, violation of the inequality is allowed. (b) $Q_{\rm min}$ (red curve) versus $\alpha$. The gray dashed line shows the bound \eqref{eq: TUR to test} to be tested (see the main text). (c)-(f) Numerical results of $\tilde{p}$, ${\rm Var}[\tilde{p}]$, $H$, and $\Delta_{\tilde{p}}$, where $\tau = \tau^*_Q(\alpha)$. The orange dashed curves show the results obtained with the large-$\alpha$ approximation.\label{fig: Q value}}
\end{figure*}

\section{Absence of a tradeoff among $p$, $\eta$, and ${\rm Var}[p]$}\label{sec: TUR}
It is proved that an inequality $p \le \Delta_p(\eta^*-\eta_{\rm cl})/\eta_{\rm cl}k_B T_{\rm cl}$ holds for CHEs \cite{Pietzonka2018universal} and CIHEs \cite{tanogami2023universal}, where $\Delta_p$ characterizes the power fluctuations and $\eta^*$ is the Carnot's efficiency for CHEs and $\eta^*=1$ for CIHEs. This tradeoff indicates that a C(I)HE operating at a finite $p$ and a finite $\eta_{\rm cl}$ cannot have vanishing power fluctuations. We report the absence of such a tradeoff relationship in our engine. 

To show the absence of the tradeoff relationship, we test the validity of the following inequality
\begin{equation}\label{eq: quantum p bound}
    p\le \frac{\Delta_p}{k_B T}\frac{1-\eta}{\eta} = \frac{t}{2}\frac{{\rm Var}[p]}{k_B T}\frac{1-\eta}{\eta},
\end{equation}
where we use the quantities defined within a single engine cycle and replace $T_{\rm cl}$ by $T$, which is the temperature of a heat bath coupled with the measurement apparatus to erase the information. The other two quantities $\Delta_p$ and ${\rm Var}[p]$ are given below.

The power fluctuation defined on a long-time run is given by $\Delta_p := \lim_{M\to +\infty} {\rm Var}[W] / 2Mt$, where ${\rm Var}[W]$ is the variance of the accumulated energy gain after time $Mt$. Because $w_i$ and $w_j$ are independent if $i\ne j$, we have $\Delta_p  = \lim_{M\to +\infty} M{\rm Var}[E]/2Mt = \lim_{M\to +\infty} M t^2{\rm Var}[p]/2Mt  = t{\rm Var}[p]/2$, where in the first equality we denote ${\rm Var}[E]$ as the variance of the energy gain for a single cycle, and in the second equality we denote 
\begin{equation}
    {\rm Var}[p] := \frac{{\rm Var}[E]}{t^2} := \frac{1}{t^2}\left[\Delta^2 \sum_0^N n^2 P(n;t) - E(t)^2\right]
\end{equation}
as the variance of the power using the definition $p := E/t$. Rearranging terms of the inequality~\eqref{eq: quantum p bound} gives an inequality $Q := {\rm Var}[p]H/p^2 \ge 2$, which is similar to the thermodynamic uncertainty relation (TUR) \cite{Barato2015thermodynamic,Liu2020thermodynamic}. Using the dimensionless quantities, we have for the TUR-like inequality
\begin{equation}\label{eq: TUR to test}
    Q(\alpha,\tau) := \frac{{\rm Var}[\tilde{p}]}{\tilde{p}^2}H \ge 2,
\end{equation}
and it can be equivalently rewritten as $\tilde{p} \le \Delta_{\tilde{p}}(1-\eta)/\tilde{T}\eta = \tau{\rm Var}[\tilde{p}](1-\eta)/2\tilde{T}\eta$, where $\Delta_{\tilde{p}} := \tau{\rm Var}[\tilde{p}]/2 = \hbar \Delta_p /J^3$ and ${\rm Var}[\tilde{p}] = \hbar^2 {\rm Var}[p]/J^4$.

We test the validity of the inequality~\eqref{eq: TUR to test} for our engine. If no violation is found, a tradeoff relationship holds for GQIEs; if it is violated but the value of $Q$ bounded from below by a different positive constant, the tradeoff relation $p\le A {\rm Var}[p](1-\eta)/\eta$ holds but with $A \ne t/2k_B T$; if the value of $Q$ can vanish, the tradeoff relation $p\le A {\rm Var}[p](1-\eta)/\eta$ does not exist.

In Fig.~\ref{fig: Q value}(a), we numerically calculate the value $Q-2$ for $\alpha\in[10^{-2},10^2]$ and $\tau\in [0,500]$. The points representing $Q-2 <0$ are plotted in red and non-negative values are plotted in gray (the larger the whiter). There exists a clear boundary located near $\alpha =3.6$ (green dashed line), to the right of which the value of $Q-2$ is always non-negative, and on the other side negative values are allowed. This means that, if $\alpha <3.6$, the TUR-like inequality \eqref{eq: TUR to test} can be violated in our engine, while for $\alpha>3.6$ the inequality holds true.

We calculate the minimum value of $Q$ for a fixed $\alpha$ defined as
\begin{equation}\label{eq: def of Q_min}
    Q_{\rm min}(\alpha) := \min_{\tau} Q(\alpha,\tau) = Q(\alpha,\tau^*_Q(\alpha)),
\end{equation}
where the ``worst" time is defined as $\tau^*_Q(\alpha) := \argmin_{\tau} Q(\alpha,\tau)$. Numerical results plotted in Fig.~\ref{fig: Q value}(b) show that $Q$ is lower-bounded by 0. Hence, a tradeoff relation of the form
\begin{equation}\label{eq: absence of a tradeoff}
    p\le A{\rm Var}[p] \frac{1-\eta}{\eta},
\end{equation}
with $A$ being any positive constant, does not hold for our engine. This implies that our engine can operate at a finite $p$ and a finite $\eta$ with vanishing ${\rm Var}[p]$, and thus can be more stable than C(I)HEs. The reason is due to the absence of thermal fluctuations and the coherent quantum evolution. This is our third main observation.

We give an explanation of the boundary near $\alpha=3.6$ by using the large-$\alpha$ approximation. From the definition of the variance of power, for a large $\alpha$ we have ${\rm Var}[\tilde{p}](\alpha,\tau) = 4\sin^2(\alpha\tau/2)/\tau^2$. Therefore, we obtain $Q(\alpha,\tau) = H(x)/x =: Q(x)$, where $x(\alpha,\tau) := 4\sin^2(\alpha\tau/2)/\alpha^2$. For $x\in (0,1)$, the function $Q(x)$ monotonically decreases with increasing $x$. Solving $Q < 2$ therefore gives $x(\alpha,\tau) > 0.31$. Because $x_{\rm max}(\alpha) := \max_{\tau} x(\alpha,\tau) = 4/\alpha^2$ which is a decreasing function of $\alpha$, we have $\alpha <3.6$, which agrees well with the numerical results.

The validity of the inequality \eqref{eq: TUR to test} for large $\alpha$ can also be understood from this analysis. In this case, $x$ is at most $ O(\alpha^{-2})$ which is very small. This implies that $Q$ has a large minimal value of $ O(2\ln\alpha)$ which agrees with the numerical results in Fig.~\ref{fig: Q value}(b). It is due to the large variance of the power in this regime as shown in Figs.~\ref{fig: Q value}(d) and (f) for ${\rm Var}[\tilde{p}]_{Q \rm min}(\alpha) := {\rm Var}[\tilde{p}](\alpha, \tau^*_Q(\alpha))$ and $\Delta_{\tilde{p}Q \rm min}(\alpha) := \Delta_{\tilde{p}}(\alpha,\tau^*_Q(\alpha))$, respectively. From the variance of power, we have $\Delta_{\tilde{p}}(\alpha,\tau) = 2\sin^2(\alpha\tau/2)/\tau$. To minimize the $Q$ value, one needs to choose $x$ as large as possible, implying that $\alpha\tau^*_Q(\alpha) = (2n-1)\pi$ for $n\in \mathbb{N}$. Because $\alpha$ is large, the minimal time $\tau^*_Q(\alpha) \approx \pi/\alpha$ is so small that ${\rm Var}[\tilde{p}]_{Q \rm min}(\alpha) \approx 4\alpha^2/ \pi^2$ and $\Delta_{\tilde{p}Q \rm min}(\alpha) \approx 2\alpha /\pi$. The power is finite and close to $\tilde{p}_{Q \rm min} \approx 4/\pi$ as shown in Fig.~\ref{fig: Q value}(c). The Shannon entropy is about $H_{Q\rm min} \approx -4/\alpha^2 \ln(4/\alpha^2)$ which is small as shown in Fig.~\ref{fig: Q value}(e) but not small enough to dominate the large fluctuations. The rapid oscillations that appear in Figs.~\ref{fig: Q value}(c), (d), and (f) are due to numerical precision in evaluating the worst time $\tau^*_Q(\alpha)$. Because of a large $\alpha$, a slight deviation from the worst time results in a significant change in the values of these quantities. Nevertheless, the profiles that bound the numerical results from above are very close to the large-$\alpha$ predictions.

We provide a qualitative understanding of the small- and the intermediate-$\alpha$ regimes. The minimal value of $Q$ almost saturates at the trivial lower bound of 0 as depicted in Fig.~\ref{fig: Q value}(b). In the small-$\alpha$ regime, due to the dominance of hopping $J$, the wave packet is expected to have a large spread which results in a large Shannon entropy as shown numerically in Fig.~\ref{fig: Q value}(e). Although the power tends to vanish as shown in Fig.~\ref{fig: maximum p and v}(a), the power fluctuation vanishes much quicker [see Figs.~\ref{fig: Q value}(d) and (f)] because $\alpha$ is so small that a large variance in the wave function cannot result in a significant variance in the power. In the intermediate regime, Bloch oscillations become relevant, and complicated interference discussed previously about the bumps again appears. The spread of the wave function is suppressed by Bloch oscillations so that a drop in the Shannon entropy arises. This constrained wave packet in turn implies a small but nonzero variance in the power.

\begin{figure*}[t!]
    \centering
    \includegraphics[width=\textwidth]{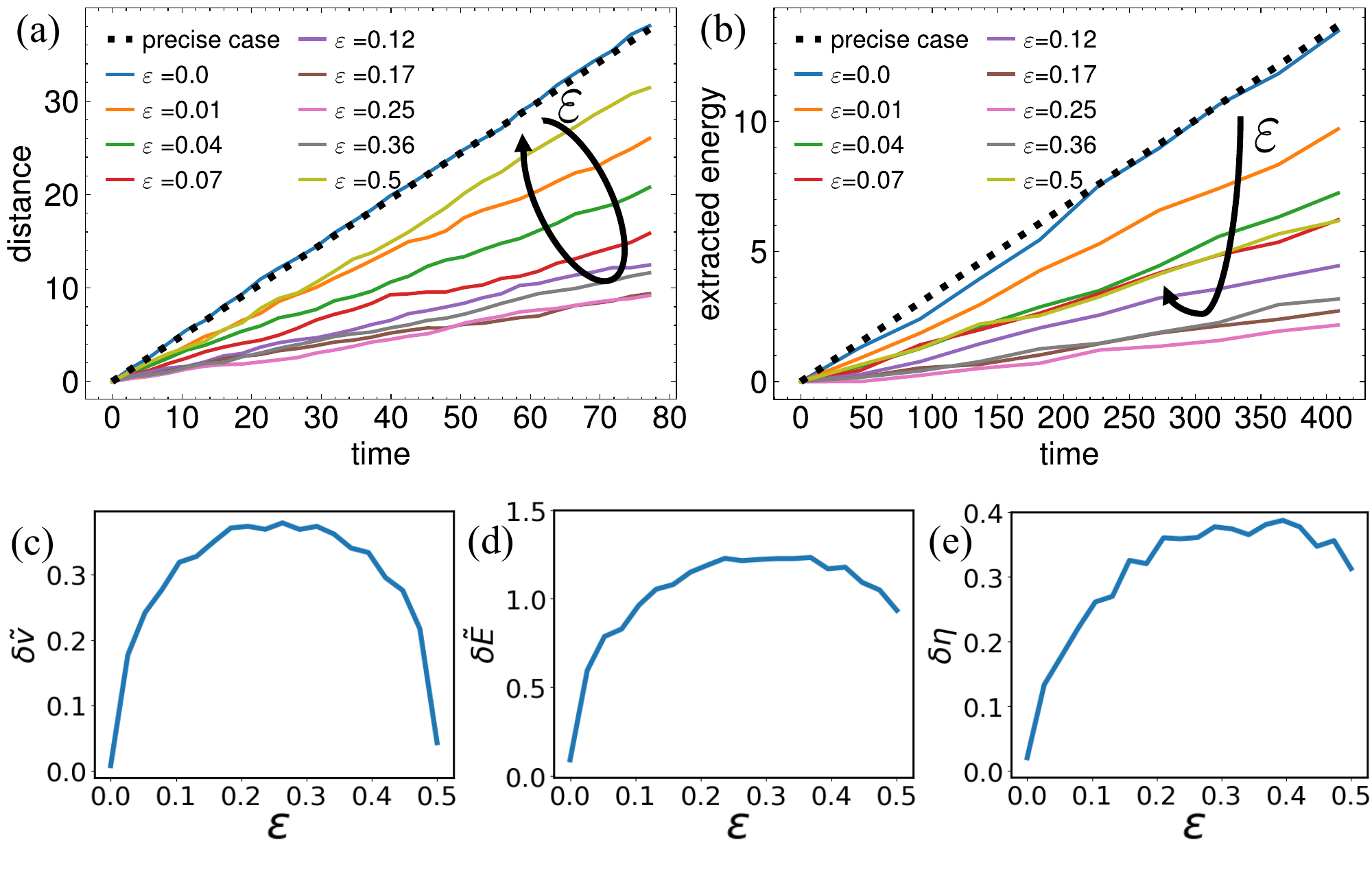}
    \caption{Influence of the error $\varepsilon$ on the performance with $\alpha =1$ for the $j-2$ FB-contolled engine. We take 9 values of $\varepsilon$ from 0 to 0.5 with a logarithmic interval as shown in the legend. We plot the trajectory ensemble average (colored curve) of (a) the traveled distance and (b) the accumulated energy gain for (a) $M=30$ cycles and (b) $M=10$ cycles where each engine cycle evolves for (a) $\tau=\tau^*_v = 2.65$ and (b) $\tau=\tau^*_E(\alpha=1) = 45.42$ for different $\varepsilon$'s. The black dotted lines show the single-cycle optimized results. (c) Change $\delta \tilde{v}$ in the average transport velocity from the error-free velocity $\tilde{v}_{\rm max}(\alpha=1) = 0.49$ as a function of $\varepsilon$. (d) Change in the average energy gain per cycle compared with the error-free value $\tilde{E}_{\rm max}(\alpha=1) = 1.52$. (e) Change in the efficiency with $\tau^*_{\eta}(\alpha=1) = 94.0$ compared with the error-free value $\eta_{\rm max} = 0.51$.\label{fig: j-2_FB_influence_by_g}}
\end{figure*}

\section{Imprecise measurements} \label{sec: impre meas}
For a quantum system with Hilbert space $\mathbf{H}$ spanned by $\{ \ket{i} \}$, a general measurement operator yielding an outcome $j$ has a general form \cite{wiseman_milburn_2009book} $M_j := \sum_i f_i(j)\Pi_i$, satisfying the completeness condition $\sum_j M_j^{\dagger}M_j=I$. A precise measurement has a set of measurement operators satisfying $\forall j$, $f_i(j)=\delta_{ij}$. The other cases are referred to as imprecise measurements. We consider general imprecise measurement. Suppose that at the beginning of a feedback cycle ($t=t_0$) the system is in state $\ket{\psi(t_0)}=\sum_n c_n(t_0)\ket{n}$, where $c_n(t_0)$'s are the coefficients. We measure the system and obtain an outcome $j$ with probability $P(j):= \bra{\psi(t_0)}M_j^{\dagger}M_j\ket{\psi(t_0)}$ and the post-measurement state becomes $\ket{\psi'(t_0)} := M_j \ket{\psi(t_0)}/\sqrt{P(j)}$ which is generally still a linear combination of all bases. Feedback control conditioned on the outcome $j$ brings the Hamiltonian from $H(t_0)$ to $H'_j(t_0)$ in an infinitesimal time as a quench \cite{Toyabe2010,Elouard2018efficient}. Due to the imprecision, this quench generally involves direct energy exchange with the feedback controller given by $W_{\rm FB}:= \bra{\psi'(t_0)}(H'_j(t_0)-H(t_0))\ket{\psi'(t_0)}$. The state evolves for time $\tau$ with this new Hamiltonian to $\ket{\psi(t_0+\tau)} = \exp(-iH'_j(t_0)\tau/\hbar)\ket{\psi'(t_0)}$. The total energy change within this cycle is $\Delta E := \bra{\psi(t_0+\tau)}H'_j(t_0)\ket{\psi(t_0+\tau)} - \bra{\psi(t_0)}H(t_0)\ket{\psi(t_0)}$ including two contributions from the direct energy exchange by applying the feedback and from the time evolution.

\begin{figure*}[t!]
    \centering
    \includegraphics[width=\textwidth]{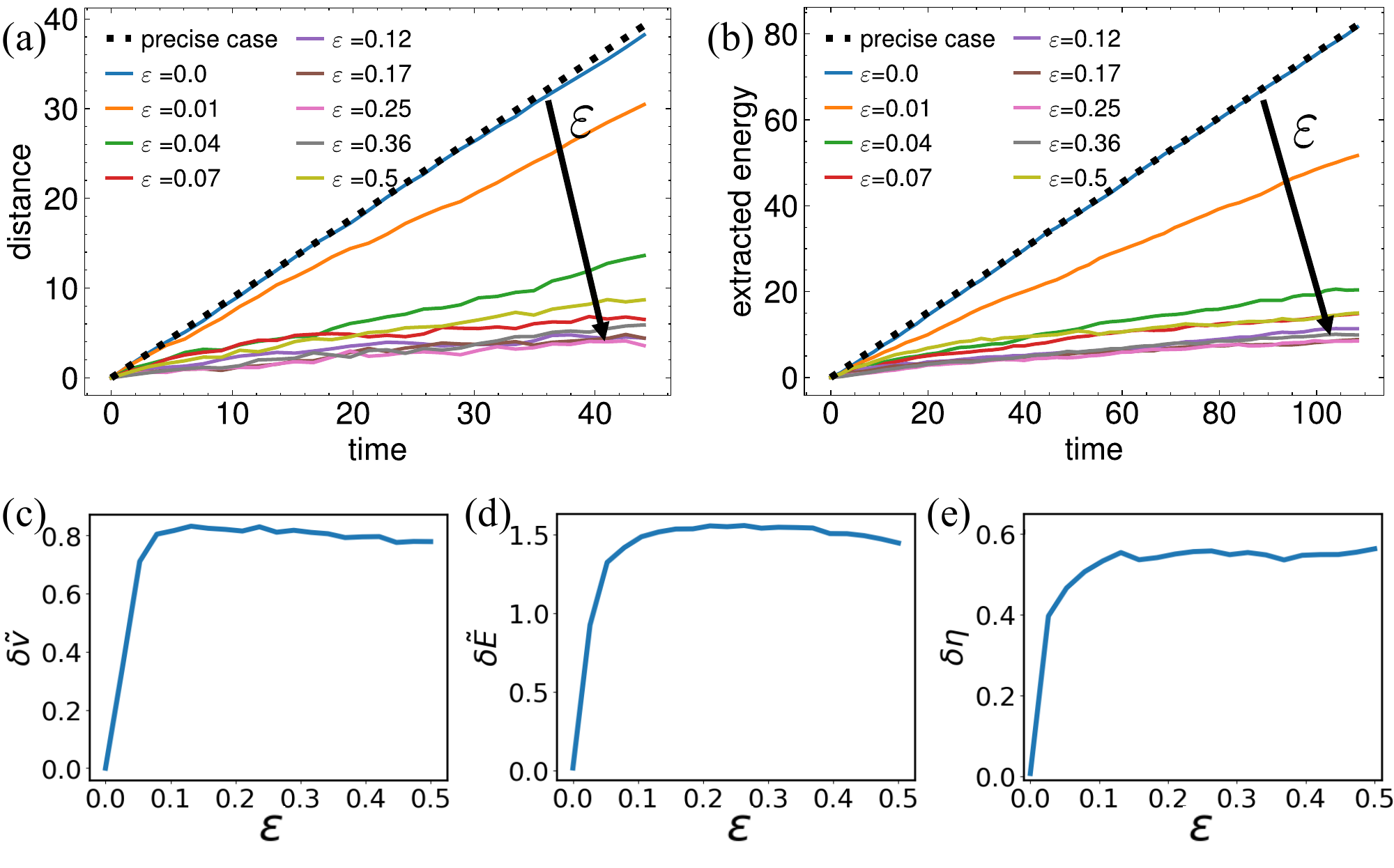}
    \caption{Influence of $\varepsilon$ on the performance with $\alpha =1$ for the $j-1$ FB-controlled engine, similarly to Fig.~\ref{fig: j-2_FB_influence_by_g} but with a difference in that we use the optimal times, velocity, and energy gain obtained in the error-free $j-1$ FB case. We use $M=30$ to plot (a) and $M=50$ to plot (b).\label{fig: weak_j-1_speed}}
\end{figure*}

To be specific in our setup, we consider a simple but realistic case in which a single-shot measurement makes an error only at nearest-neighbor sites \cite{Gross2021quantum}:
\begin{align}\label{eq: def of our imprecise meas op}
    M_j := \begin{cases}
        \sqrt{1-\varepsilon}\Pi_{j} + \sqrt{\varepsilon} \Pi_{j+1}, & j = 0;\\
        \sqrt{1-2\varepsilon}\Pi_{j} + \sqrt{\varepsilon} \left(\Pi_{j-1} + \Pi_{j+1} \right), & 0 < j < N;\\
        \sqrt{1-\varepsilon}\Pi_{j} + \sqrt{\varepsilon} \Pi_{j-1}, & j = N.
    \end{cases}
\end{align}
This set of measurement operators means that if the measurement outcome $j$ is at neither of the two end sites $n =0$ and $n = N$, a $100\varepsilon \%$ error is made at either of the nearest-neighbor sites $n = j-1$ and $n = j+1$. For an outcome $j = 0$ or $j = N$, a $100\varepsilon \%$ error only occurs at one site (i.e., $j=1$ or $j=N-1$). To fulfill the completeness condition for the measurement operators, the coefficient of $\Pi_0$ or $\Pi_N$ is modified to $\sqrt{1-\varepsilon}$. For $\varepsilon = 0$, we recover the precise position measurement $M_j = \Pi_j$. For $\varepsilon = 0.5$, we make the largest ($50 \%$ at each neighbor site) error in evaluating the position of the particle. We assume that the value of $\varepsilon$ is known, which is a common assumption in quantum control \cite{Fuchs2001information,Brannczyk2007quantum,Xiao2011reexamination}. Because the state-of-the-art experimental techniques can achieve a level of imprecision as low as $\varepsilon\lesssim$ (several) \%  when local control is performed \cite{Gross2021quantum}, we are mostly interested in a low error level $\varepsilon\lesssim 5\%$ while still presenting numerical results over the whole range of $\varepsilon$.

Suppose that the pre-measurement state is a general pure state given by $\ket{\psi} = \sum_n c_n \ket{n}$, with $\sum_n |c_n|^2 = 1$. The probability of obtaining a measurement outcome $j$ is given by $P(j)=\bra{\psi}M_j^{\dagger}M_j\ket{\psi}$, and the post-measurement state with a measurement outcome $0<j<N$ is given by
\begin{align}\label{eq: post-meas state with imprecision}
    \ket{\psi'} & = \frac{M_j \ket{\psi}}{\sqrt{P(j)}}  \nonumber\\
    & = \frac{\sqrt{1-2\varepsilon}}{\sqrt{P(j)}}c_j \ket{j} + \frac{\sqrt{\varepsilon}}{\sqrt{P(j)}} \left( c_{j-1} \ket{j-1} + c_{j+1} \ket{j+1} \right),
\end{align}
which is still a pure state but in a superposition. Thus, with measurement operators given in Eq.~\eqref{eq: def of our imprecise meas op}, the purity of the state of the measured system is preserved along each trajectory. In the numerical simulation, we choose the ground state of the initial Hamiltonian $H_0$ to be the initial state of the first cycle.

With a measurement outcome $j \ne 0$, the post-measurement state always has a nonzero probability amplitude $\sqrt{\varepsilon}c_{j-1}/\sqrt{P(j)}$ at the left neighbor site $j-1$ for a nonzero $\varepsilon$. This means that if we apply a potential $V$ at site $j-1$, an amount of work $\varepsilon|c_{j-1}|^2 V/P(j)$ must be applied to quench the Hamiltonian to $H_j = H_{\rm WS} + V(\varepsilon)\ket{j-1}\bra{j-1}$ with a finite $V(\varepsilon)$. For the sake of clarity, we first consider a slightly different FB to circumvent this extra work injection. If the measurement outcome is $j \ge 2$, a sufficiently large potential $V \gg \max{\{ J, \Delta\} }$ is suddenly applied at site $j-2$, i.e., the left next-nearest-neighbor site; if the measurement outcome is $j = 0$ or $1$, the left boundary is kept at $-1$ because we simulate the dynamics with a finite chain from $n = 0$ to $n = N$. We call this FB protocol the $j-2$ FB. The absence of work injections by FB makes the sole effects caused by the presence of measurement imprecision clearer. On the other hand, applying a potential wall at site $j-1$ while obtaining an outcome $j$ will be referred to as the $j-1$ FB. For both FB protocols, we perform extensive numerical simulations to show that our engine still gains work cumulatively and transports the particle unidirectionally more slowly than the precise-measurement case. 

\subsection{Performance for the $j-2$ FB protocol}
We study the influence of measurement errors on the performance of our engine with the $j-2$ FB protocol. Due to the failure of single-cycle calculations, it is difficult to optimize the performance to obtain functions such as $\tilde{p}_{\rm max}(\alpha,\varepsilon)$ and $\tau^*(\alpha,\varepsilon)$ with respect to $\alpha$ and $\varepsilon$. To study the sole influence of measurement imprecision, we realize many long trajectories and take ensemble averages of the total energy gain $\tilde{E}$, power $\tilde{p}$, and efficiency $\eta$ while keeping $\alpha=1$ held fixed and setting the evolution time for each cycle to be the optimal one $\tau^*_{E/p/\eta}(\alpha=1)$ that is obtained in precise measurement optimization. The applied potential barrier is set to be $\tilde{V} = 10^3$. The results are presented in Fig.~\ref{fig: j-2_FB_influence_by_g}.

In Fig.~\ref{fig: j-2_FB_influence_by_g}(a), we show the influence of $\varepsilon$ on the distance traveled by the particle. The operation time for each cycle is chosen to be $\tau^*_v = \tau^*_p(\alpha=1) = 2.65$ obtained in a similar single-cycle optimization of the $j-2$ FB with the precise measurement as done in Fig.~\ref{fig: maximum p and v} for the $j-1$ FB with precise measurements. The single-cycle optimized velocity for $\varepsilon=0$ gives the black dotted line representing $n(m\tau^*_v) = \tilde{v}_{\rm max}(\alpha=1) m\tau^*_v = 0.49 m\tau^*_v$. For each out of 9 values of $\varepsilon$ varying from $\varepsilon=0$ to $\varepsilon=0.5$ with a logarithmic interval, we run 100 realizations, each of which consists of $M=30$ cycles. Each colored curve represents the ensemble average of the traveled distance for a given value of $\varepsilon$. We find that the presence of $\varepsilon$ decreases the transported distance after the same number of cycles. The average velocity for each colored curve is the ratio of the final traveled distance to the total time of a trajectory. We plot in Fig.~\ref{fig: j-2_FB_influence_by_g}(c) the change $\delta \tilde{v}$ in the average velocity compared with the velocity $\tilde{v}_{\rm max}(\alpha=1) = 0.49$ in the error-free case. The velocity is less than the error-free velocity with $\varepsilon>0$ but remains positive and finite.

We notice a recovery behavior of the mean velocity as the imprecision $\epsilon$ becomes large as shown in Fig.~\ref{fig: j-2_FB_influence_by_g}(c). This can be understood from the measurement operator \eqref{eq: def of our imprecise meas op}. In the maximally imprecise limit $\epsilon\to 0.5$, the coefficient $\sqrt{1-2\epsilon}$ of $\Pi_j$ vanishes, so the post-measurement state conditioned on outcome $j$ has a negligible weight on $|j\rangle$ and is supported mainly on $|j\pm 1\rangle$. The hard wall placed at $j-2$ makes $j-1$ the boundary of the semi-infinite chain. A finite population is prepared at site $j-1$, which enhances the rectification of the positional information and leads to the observed recovery at a large $\epsilon$.

We also show the total accumulated energy gain in Fig.~\ref{fig: j-2_FB_influence_by_g}(b). We take the evolution time as $\tau^*_E(\alpha=1) = 45.42$ and $M=10$ in each trajectory. The black dotted line is taken from the single-cycle optimization with $\tilde{E}_{\rm max}(\alpha=1) = 1.52$. We observe reduced but still positive accumulated energies with a nonzero $\varepsilon$. In Fig.~\ref{fig: j-2_FB_influence_by_g}(d), we plot the reduction in the average energy gain per cycle as a function of $\varepsilon$. It exhibits behavior similar to the difference in the mean velocity. 

We finally show the reduction in the efficiency in Fig.~\ref{fig: j-2_FB_influence_by_g}(e). The efficiency in the error-free case is obtained by the single-cycle optimization and given by $\eta_{\rm max}(1) = 0.51$. The corresponding optimal time is $\tau^*_{\eta}(\alpha=1) = 94.0$. The behavior of the efficiency is similar to that of the energy gain per cycle.

These results suggest that with the measurement error, the performance of our engine typically deteriorates. This is reasonable because an erroneous measurement cannot extract all pieces of information contained in the coherence or the superposition of the quantum pure state. Despite the deterioration in performance, our engine can still gain a positive amount of energy in its potential energy and also transport the particle uphill against the linear potential, even if we make a large error in the measurements. Especially, if the error level can be constrained within a realistic level of about 5\% \cite{Preiss2015strongly,Gross2021quantum}, a reasonable performance can be ensured, implying the robustness of our engine.

\subsection{Performance for the $j-1$ FB protocol}\label{app:j-1 performance}
We first introduce the wall height $\tilde{V}(\varepsilon)$ used in the $j-1$ FB in the presence of measurement errors and define the direct work injection $\tilde{E}_{\rm FB}$ by applying FB in one cycle. Let us consider the $m$th engine cycle. According to the discussion about Eq.~\eqref{eq: post-meas state with imprecision}, a FB applied at site $j_m -1$ for outcome $j_m$ requires direct work injection by the amount
\begin{align}\label{eq: def of stochastic E_FB}
        \tilde{E}_{\rm FB}(m) = \frac{\tilde{V}(\varepsilon)\varepsilon |c_{j_m -1}|^2}{P(j_m)}.
\end{align} 
On the other hand, the total energy gained by the particle in the $m$th engine cycle is given by the change in its internal energy during this cycle:
\begin{align}
    \tilde{w}_m := & \bra{\psi(m\tau)}\tilde{H}_{j_m}\ket{\psi(m\tau)}\nonumber\\
    & - \bra{\psi((m-1)\tau)}\tilde{H}_{j_{m-1}}\ket{\psi((m-1)\tau)},
\end{align}
where $\psi(m\tau):= \exp(-i\tilde{H}_{j_m}\tau)\ket{\psi_{m}}$ is the time-evolved state of the $m$th engine cycle. We note that the infinite-wall approximation of the quenched Hamiltonian given by Eq.~\eqref{eq: our limiting H} is no longer applicable. The total energy gain $\tilde{w}_m$ for each cycle is a stochastic variable with an average of order $ O(1)$ for precise measurements with $\alpha=1$. Therefore, we choose 
\begin{align}\label{eq:tilde v}
    \tilde{V}(\varepsilon) := \frac{0.1}{\varepsilon}
\end{align}
to yield a small work injection compared with the total energy gain which includes both the FB-injected work and the energy extracted from quantum fluctuations. Summing $\tilde{w}_m$ for all cycles $m\in[0, M]$ and taking the average over many trajectories, we have the average total accumulated energy $\tilde{W}$ consisting of (1) the average FB-injected energy $\tilde{W}_{\rm FB} := \sum_m \tilde{E}_{\rm FB}(m) /M$ and (2) the energy $\tilde{W}_{\rm ext.meas} := \tilde{W} -\tilde{W}_{\rm FB} $ extracted from measurements. The relative contribution of these two parts is numerically investigated and the results are presented in Fig.~\ref{fig: weak_j-1_energy} in Appendix~\ref{app:FB work}.

We investigate the influence due to a finite error on the performance of our engine. We perform similar numerical simulations as done for the $j-2$ FB. In Fig.~\ref{fig: weak_j-1_speed}, we present the results of our numerical study on the traveled distance (velocity $\tilde{v}$), the total accumulated energy $\tilde{W}$, and the efficiency $\eta$ for 9 different values of $\varepsilon$ ranging logarithmically from 0 to 0.5.

We plot the average traveled distance in Fig.~\ref{fig: weak_j-1_speed}(a). Each colored solid line is plotted by taking the numerical average over 100 trajectories for $M= 30$ engine cycle per trajectory with a specific value of $\varepsilon$. All the calculations are performed with $\tau^*_v(1) = 1.52$. The average velocity in the error-free case is $\tilde{v}_{\rm max}(1) = 0.89$ which gives the black dotted line. As the error level increases, the mean velocity drops rapidly to a positive but small value of about 0.1 if the error is larger than 10\%. We also calculate the average total accumulated energy $\tilde{W}$ in Fig.~\ref{fig: weak_j-1_speed}(b) and the average total energy gain per cycle $\tilde{E}$ in (d), respectively. The evolution time per cycle is $\tau^*_E(\alpha=1) = 2.21$ for $M=50$ cycles. The energy gain per cycle in the error-free case is $\tilde{E}_{\rm max}(\alpha=1) = 1.68$. A similar behavior to the mean velocity in Figs.~\ref{fig: weak_j-1_speed}(a) and (c) is observed. The recovery behavior of the mean velocity observed in the $j-2$ FB protocol [Fig.~\ref{fig: j-2_FB_influence_by_g}(c)] is absent for the $j-1$ FB protocol as shown in Fig.~\ref{fig: weak_j-1_speed}(c). The reason is that we choose $V(\epsilon)=0.1J/\epsilon$ to suppress the injected work. The barrier height becomes as low as $V(0.5)=0.2J$, thus, it cannot efficiently prevent backward motion and no recovery is observed.

Finally, we study the efficiency of our engine. We define the efficiency as the ratio of total work gain to the total energy input. Here the total energy gain is $\tilde{W}$ for $M$ cycles. The total energy input $\tilde{W}_{\rm total}$ consists of three contributions. The first one is the energy $\tilde{W}_{\rm meas}$ required to perform quantum measurements; the second one is the work $\tilde{W}_{\rm FB}$ injected through application of the FB protocol; the third one comes from the energy $\tilde{W}_{\rm eras}$ to erase the information stored in the measurement apparatus after each cycle. Therefore, the total energy input is given by
\begin{align}
    \tilde{W}_{\rm total}  & := \tilde{W}_{\rm meas} +\tilde{W}_{\rm FB} + \tilde{W}_{\rm eras} \nonumber\\
    & = \tilde{W}_{\rm ext.meas} + \tilde{T} \sum_{m=0}^M H(m) + \tilde{W}_{\rm FB}  \nonumber\\
    & = \tilde{W} + \tilde{T} \sum_{m=0}^M H(m),
\end{align}
where we have $\tilde{W}_{\rm meas}+\tilde{W}_{\rm eras} = \tilde{W}_{\rm ext.meas} + \tilde{T} \sum_{m=0}^M H(m)$ and $H(m) := -\sum_n P(n,m\tau) \ln P(n,m\tau)$ is the Shannon entropy of the measurement outcomes in the $m$th cycle \cite{Sagawa2009minimal,Sagawa2011erratum,Abdelkhalek2016quantum}. We assume that the structure of our measurement apparatus is simple enough to have the relation $S\left(\sum_n P(n,m\tau) \rho_n (m\tau)\right) - S\left(\rho_{0,\rm can}\right) = H(m)$, $\forall m$, where $\rho_n(m\tau)$ is the density operator of the measurement apparatus at the end of the $m$th engine cycle associated with an outcome $n$, and $\rho_{0,\rm can}$ is a default canonical state of the measurement apparatus to which the probe is reset after every cycle. Therefore, the efficiency of our engine is given by
\begin{equation}
    \eta(\varepsilon) := \lim_{M\to \infty}\frac{\tilde{W}}{ \tilde{W}_{\rm total}} = \lim_{M\to \infty}\frac{\tilde{W}}{\tilde{W} + \tilde{T} \sum_{m=0}^M H(m)},
\end{equation}
where a finite error $\varepsilon$ affects the values of both the total energy gain $\tilde{W}$ and the Shannon entropy $\sum_{m=0}^M H(m)$. We plot the difference in the efficiency defined as $\delta \eta := \eta_{\rm max}(\alpha=1) -  \eta(\varepsilon)$ as a function of $\varepsilon$ in Fig.~\ref{fig: weak_j-1_speed}(e) with $\tau^*_{\eta}(\alpha=1) = 5.94$ for $M=50$ cycles and the efficiency $\eta_{\rm max}(1) = 0.61$ in the error-free case. We again observe a behavior similar to the mean velocity in Fig.~\ref{fig: weak_j-1_speed}(c) and the mean energy gain per cycle in Fig.~\ref{fig: weak_j-1_speed}(d).

The numerical results imply some perspectives of our engine with measurement imprecision. First, a finite measurement imprecision $\varepsilon$ is generally harmful to the performance of our engine. Second, even with imprecision, our engine still works to accumulate energy and transport a particle, while consuming direct work injected by applying FB. Third, the performance drops quickly with increasing imprecision. While the level of imprecision is lower than 5\%, one still ensures a reasonable performance and the engine operates with little direct energy injection. This can be understood intuitively from the fact that the potential height of the wall $\tilde{V}(\varepsilon) = 0.1 / \varepsilon$ with $\varepsilon = 0.05$ becomes comparable to the potential step height $\alpha = 1$ in this case, indicating that the particle very likely hops to the left for $\tilde{V}(\varepsilon) - \alpha \sim \alpha$. This also implies that the allowed error $\varepsilon$ to maintain the performance depends on the slope $\alpha$.

\section{Experimental implementation} \label{sec: exp}
In our engine, every step in the engine cycle can be implemented with existing cold-atom experimental techniques.
The 1D tilted periodic lattice with Hamiltonian $H_{\rm WS}$ \eqref{eq: WS Hamiltonian} has been widely implemented and utilized in experiments of cold atoms in optical lattices \cite{Simon2011quantum,Fukuhara2013quantum,Preiss2015strongly,Scherg2021observing,Natale2022Bloch} and many other systems \cite{Mendez1988stark,Voisin1988observation,Waschke1993coherent,Unuma2023effects,Morsch2001Bloch,Pertsch2002anomalous}. Isolated ultracold atomic systems offer an ideal platform to study genuinely quantum information engines because the effect of an environment can be almost neglected and a high precision measurement and control of the quantum system is possible. We describe how to realize our engine by using an ultracold atom in optical lattices.

In Ref.~\cite{Preiss2015strongly}, ultracold atoms of bosonic $^{87}$Rb are loaded into optical lattices to study quantum walk. In this optical lattice, the Hamiltonian is given by the Bose-Hubbard Hamiltonian
\begin{equation}\label{eq: Bose Hubbard model}
    H_{\rm BH} = -J\sum_{\langle i,j \rangle}a_i^{\dagger}a_j + \frac{U}{2}\sum_i n_i (n_i -1) + \Delta \sum_i i n_i,
\end{equation}
where $a_i^{\dagger}$ and $a_i$ are the bosonic creation and annihilation operators, respectively, and $n_i := a_i^{\dagger}a_i$ is the number operator on site $i$, $J$ is the hopping amplitude, $U$ is the strength of a repulsive on-site interaction, and $\Delta$ is the strength of an external linear field. The values of $J$ and $U$ can be tuned by controlling the depth of the optical lattice. The external linear field is implemented with a magnetic field gradient. The optical lattice used in Ref.~\cite{Preiss2015strongly} is two-dimensional, while each horizontal tube in the $x$-axis is decoupled from each other. With this 2D lattice consisting of decoupled 1D tubes, a large number of simultaneous realizations is possible.

Initially, in every horizontal tube, a single atom is prepared in a state localized at site $0$ by the single-site addressing technique \cite{Weitenberg2011addressing,Viscor2012single,Fukuhara2013quantum,Wang2015coherent,Zupancic2016ultra,Koepsell2019imaging,Ji2021coupling} with high fidelity. The dynamics of the positions of the atoms is recorded with single-site resolution using fluorescence imaging in a deep optical lattice with the technique of quantum-gas microscopy \cite{Bakr2009quantum,Sherson2010single,Haller2015single,Cheuk2015quantum,Parsons2015site,Yamamoto2017site,Gross2017quantum,Okuno2020schemes,Yamamoto2020single,Schafer2020}. For each individual realization, the particle is detected on a single lattice site with high precision. Taking the average over many realizations gives the probability distribution of a single particle.

A concrete example of measurement imprecision made at nearest-neighbor sites can also be found in Ref.~\cite{Preiss2015strongly}. Quantitatively, the state-of-the-art experimental techniques can achieve the fidelity which is well above 99\% to distinguish between zero and one atom on a site in quantum gas microscopy \cite{Gross2021quantum}, corresponding to the error level of about $\varepsilon \lesssim 0.005$ in our analyses. However, if one wants to perform FB, one might have to reduce the number of photons for imaging to mitigate photon recoil \cite{Gross2021quantum}. In that case, the precision will be reduced and the analyses about imprecision become even more relevant.

Here the FB operation is to place a large potential at one site according to the observed position of the particle. This can be completed again with the help of single-site addressing \cite{Weitenberg2011addressing,Viscor2012single,Fukuhara2013quantum,Wang2015coherent,Zupancic2016ultra,Koepsell2019imaging,Ji2021coupling}. What one should be careful of is the time interval to apply FB. In the experimental setup used in Ref.~\cite{Preiss2015strongly}, the hopping amplitude $J$ has an order of 100Hz $hbar$. The strength $\Delta = \alpha J$ of the linear field is tunable. The timescale for applying FB should be much smaller than that of the dynamics under the Hamiltonian \eqref{eq: Bose Hubbard model}. Therefore, if $J$ is of the order of 100Hz $\hbar$, it is ideal to have $\tau_{\rm FB} \ll \min\{\hbar/J, \hbar/\Delta \} = \min\{0.01, 0.01/\alpha\}$ seconds. We notice that the optimal time for attaining the best performance of our engine is basically of the order of $0.01/\alpha$ seconds. This requirement is not so stringent because it is possible to change the depth of the optical lattice to suppress the hopping (that is to lower $J$) during placing the feedback potential. Therefore, the timescale has an order of milliseconds. While $\alpha$ is close to or much smaller than 1, the requirement on the timescale of FB is less stringent, and it is given by $\tau_{\rm FB} \ll 10$ms; if $\alpha$ is much larger than 1, the requirement can be ensured by either performing a faster FB or by changing the depth of the lattice to lower $J$, both of which are possible with current techniques.

\section{Discussion and outlook}\label{sec: discussion}
We have constructed a GQIE that rectifies only quantum fluctuations. It has well-defined performances and exhibits many interesting features including a tradeoff between the maximal power and the maximal velocity, a unit optimal efficiency, and an absence of a tradeoff of power, power variance, and efficiency which is present for the C(I)HE. The influence of measurement imprecision is also examined. Possible implementation with existing cold-atom experimental techniques and the requirements for the time scale of applying FB are discussed. 

An important question is that although the minimal energy cost $E_{\rm cost} = E + k_B T H$ is correct, it can only be achieved approximately unless an infinite amount of resources or complexity is available. Different types of nontrivial additional costs are required to be more realistic \cite{Henrik2017third,tajima2019coherencevariance,Tajima2020coherence,Guryanova2020idealprojective,Zhen2021universal,Lee2022speed,Scandi2022minimally,Taranto2023landauer}. It is of practical significance to calculate the efficiency by explicitly considering those thermodynamic costs.

\begin{acknowledgments}
We acknowledge David A. Sivak, Shoki Sugimoto, Koki Shiraishi, Zongping Gong, Hongchao Li, Philip Taranto, Mio Murao, Sosuke Ito, Naomichi Hatano, Chikara Furusawa,Yasushi Okada, Jie Gu, and Takeshi Fukuhara for fruitful discussions. This work was supported by the Scientific Research Start-up Foundation of Xihua University (Grant No: Z241064) and KAKENHI Grant No. JP22H01152 from the Japan Society for the Promotion of Science (JSPS). K.L. acknowledges support from the Global Science Graduate Course (GSGC) program of the University of Tokyo and JSPS KAKENHI Grant No. JP23KJ0637. M.N. was supported by JSPS KAKENHI Grant No. JP20K14383 and No. JP24K16989. We gratefully acknowledge the support from the CREST program ``Quantum Frontiers" (Grant No. JPMJCR23I1) by the Japan Science and Technology Agency.
\end{acknowledgments}

\appendix
\setcounter{equation}{0}
\setcounter{figure}{0}
\renewcommand{\theequation}{S\arabic{equation}}
\renewcommand{\thefigure}{S\arabic{figure}}
\section{Details for precise measurements}

\subsection{Justification of the effectively truncated semi-infinite chain with a large potential} \label{app:large V truncation}
In Step 2 of the main text, we implement the feedback control by applying a large on-site potential $V$ to site $j-1$ given a measurement outcome of $j$. For $V\gg \Delta, J$, the effective Hamiltonian is a semi-infinite tilted chain given by Eq. \eqref{eq: our limiting H} with a hard wall imposed at site $j-1$. This approximation is valid as long as leakage due to quantum tunneling of the wave function to the blocked half-space $n\le j-2$ is negligible during the unitary evolution in Step 3. We here justfity this assumption by analyzing the scaling of the leakage probability with respect to $\tilde{V}=V/J$. Qualitatively, because the on-site potential is a lattice version of a delta-function potential in the conventional quantum tunneling problem, the probability of leakage decreases with an approximately inverse-square scaling with increasing $\tilde{V}$. Below we provide a quantitative analysis.

We assume that a large potential $\tilde{V}$ is applied at site $n=0$ and that the initial state is localized at site $n=1$, i.e., \ $\ket{\psi(0)}=\ket{1}$. The probability of the leakage to the sites $n\le -1$ at time $\tau$ is
\begin{equation}
P_{\rm leak}(\tau)  := \sum_{n\le -1}\left|\bra{n}e^{-i\tilde{H}_1\tau}\ket{1}\right|^2.
\label{eq:P_leak_def}
\end{equation}

While a direct analysis of $P_{\rm leak}(\tau)$ is unfeasible, we can simplify the problem by focusing on the local three-site subspace spanned by $\{\ket{1},\ket{0},\ket{-1}\}$.
Within this subspace, the on-site energies are
\begin{equation}
E_{1}=\alpha,\quad  E_{0}=\tilde{V},\quad  E_{-1}=-\alpha,
\label{eq:E_levels}
\end{equation}
and the nonzero hopping matrix elements connecting these sites are
\begin{align}
& \bra{1}\tilde{H}_1\ket{0}=\bra{0}\tilde{H}_1\ket{1}=-1,\nonumber \\
& \bra{-1}\tilde{H}_1\ket{0}=\bra{0}\tilde{H}_1\ket{-1}=-1.
\label{eq:hopping_elements}
\end{align}
There is no direct hopping between $\ket{1}$ and $\ket{-1}$; hence, any effective coupling between these two sites is mediated by $\ket{0}$.

We define projectors $\mathbb{P} := \ket{1}\bra{1}+\ket{-1}\bra{-1}$ and $\mathbb{Q} :=\ket{0}\bra{0}$.
The Hamiltonian in this local three-site sector can be written as a block matrix with respect to $\mathbb{P}\oplus\mathbb{Q}$:
\begin{equation}
\tilde{H}_1 =
\begin{pmatrix}
\mathbb{P}\tilde{H}_1\mathbb{P} & \mathbb{P}\tilde{H}_1\mathbb{Q} \\
\mathbb{Q}\tilde{H}_1\mathbb{P} & \mathbb{Q}\tilde{H}_1\mathbb{Q}
\end{pmatrix} := \begin{pmatrix}
H_{PP} & H_{PQ} \\
H_{QP} & H_{QQ}
\end{pmatrix}
\end{equation}
with
\begin{align}
& H_{PP} =
\begin{pmatrix}
\alpha & 0 \\
0 & -\alpha
\end{pmatrix},\quad 
H_{PQ} =
\begin{pmatrix}
-1 \\
-1
\end{pmatrix},\nonumber \\
&H_{QP} =
\begin{pmatrix}
-1 & -1
\end{pmatrix},\quad 
H_{QQ} =
(\tilde{V}).
\end{align}

\begin{figure}[t!]
\begin{center}
\includegraphics[width=\columnwidth]{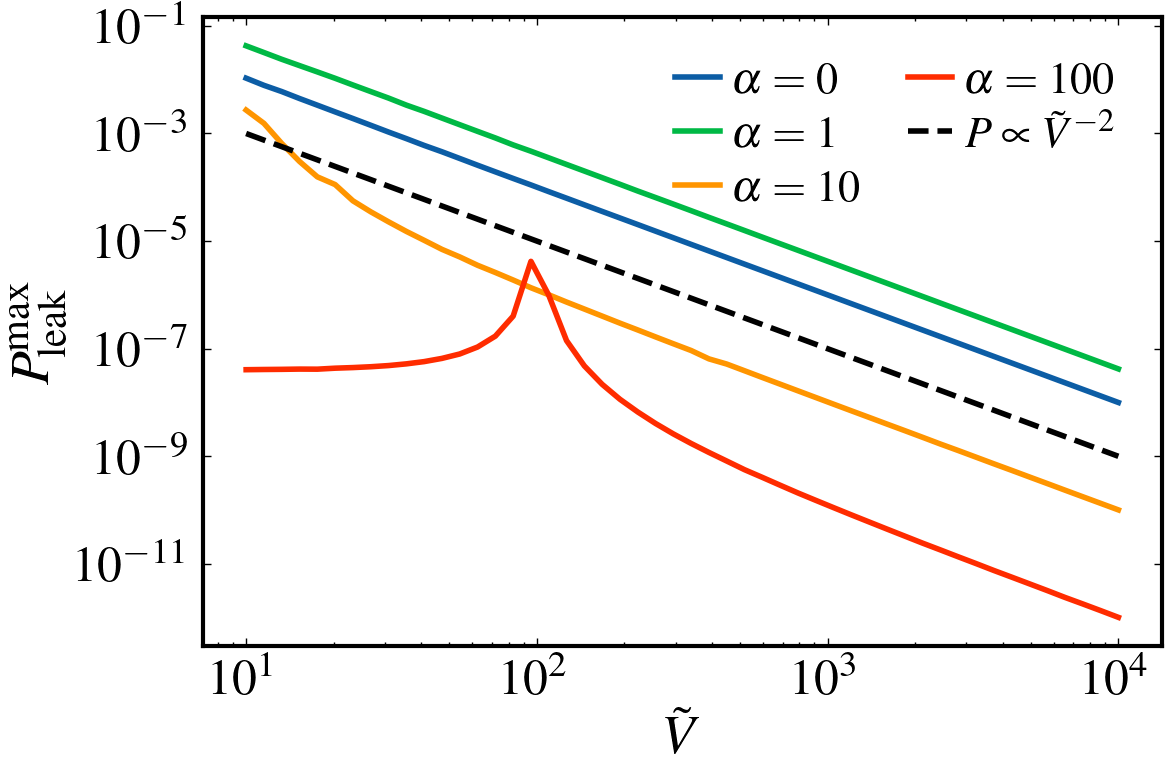}
\end{center}
\caption{Log-log plot of the maximal leakage probability $P_{\rm leak}^{\max}:=\max_{\tau} P_{\rm leak}(\tau)$ for $\tau\in (0,50]$ as a function of the barrier height $\tilde{V}$ for different values of $\alpha$. The dashed line shows the scaling relation $P_{\rm leak}\sim \tilde{V}^{-2}$.}
\label{fig: leakage scaling}
\end{figure}

The standard second-order Schrieffer-Wolff transformation \cite{Schrieffer1966relation} to eliminate the high-energy barrier site $\ket{0}$ perturbatively in $1/\tilde{V}$ yields an effective coupling between $\ket{1}$ and $\ket{-1}$ as
\begin{align}
    J_{ij} = & (H_{PP})_{ij} \nonumber\\
    & + \frac{1}{2}(\tilde{H}_1)_{i0} (\tilde{H}_1)_{0j}\left( \frac{1}{E_i - E_0} + \frac{1}{E_j - E_0} \right),
\end{align}
where $i, j \in \{+1, -1\}$ and $E_i$'s are the on-site energies given in Eq.~\eqref{eq:E_levels}. Using Eqs.~\eqref{eq:E_levels} and \eqref{eq:hopping_elements}, we obtain
\begin{equation}
    J_{1,-1} = \frac{1}{2}\left(\frac{1}{\alpha-\tilde{V}} + \frac{1}{-\alpha-\tilde{V}}\right) = -\frac{\tilde{V}}{\tilde{V}^2 - \alpha^2}.
\end{equation}
The leakage probability is then approximately proportional to $|J_{1,-1}|^2$, yielding
\begin{equation}
P_{\rm leak}(\tau) \sim \frac{\tilde{V}^2}{(\tilde{V}^2 - \alpha^2)^2} \implies P_{\rm leak}(\tau) \sim \frac{1}{\tilde{V}^2} \  \text{as} \  \tilde{V}\to \infty.
\label{eq:P_leak_scaling}
\end{equation}
This scaling relation is numerically verified by calculating the exact leakage probability defined in Eq.~\eqref{eq:P_leak_def}. In Fig.~\ref{fig: leakage scaling}, we plot $P_{\rm leak}^{\max}:=\max_{\tau} P_{\rm leak}(\tau)$ for $\tau\in (0,\tau_{\rm max}]$ as a function of $\tilde{V}$ for $\alpha \in \{0,1,10,100\}$ with $\tau_{\mathrm{max}}=50$ and $N=10^3 \gg 2\tau_{\rm max}$ to avoid boundary reflection. The numerical results agree well with the inverse-square scaling relation \eqref{eq:P_leak_scaling} for $\tilde{V}\gg \max\{1, \alpha\}$ for all values of $\alpha$.

We also mention that this approximation is no longer valid when $\tilde{V}\approx \alpha$ in the resonance regime. In this case, the leakage probability is no longer suppressed and the effective semi-infinite chain description breaks down. This behavior is clearly observed in the red curve in Fig.~\ref{fig: leakage scaling} at $\tilde{V}=10^2$ and discussed in the last part of Sec.~\ref{app:j-1 performance} for the $j-1$ FB case.

\begin{figure}[t!]
    \centering
    \includegraphics[width=\columnwidth]{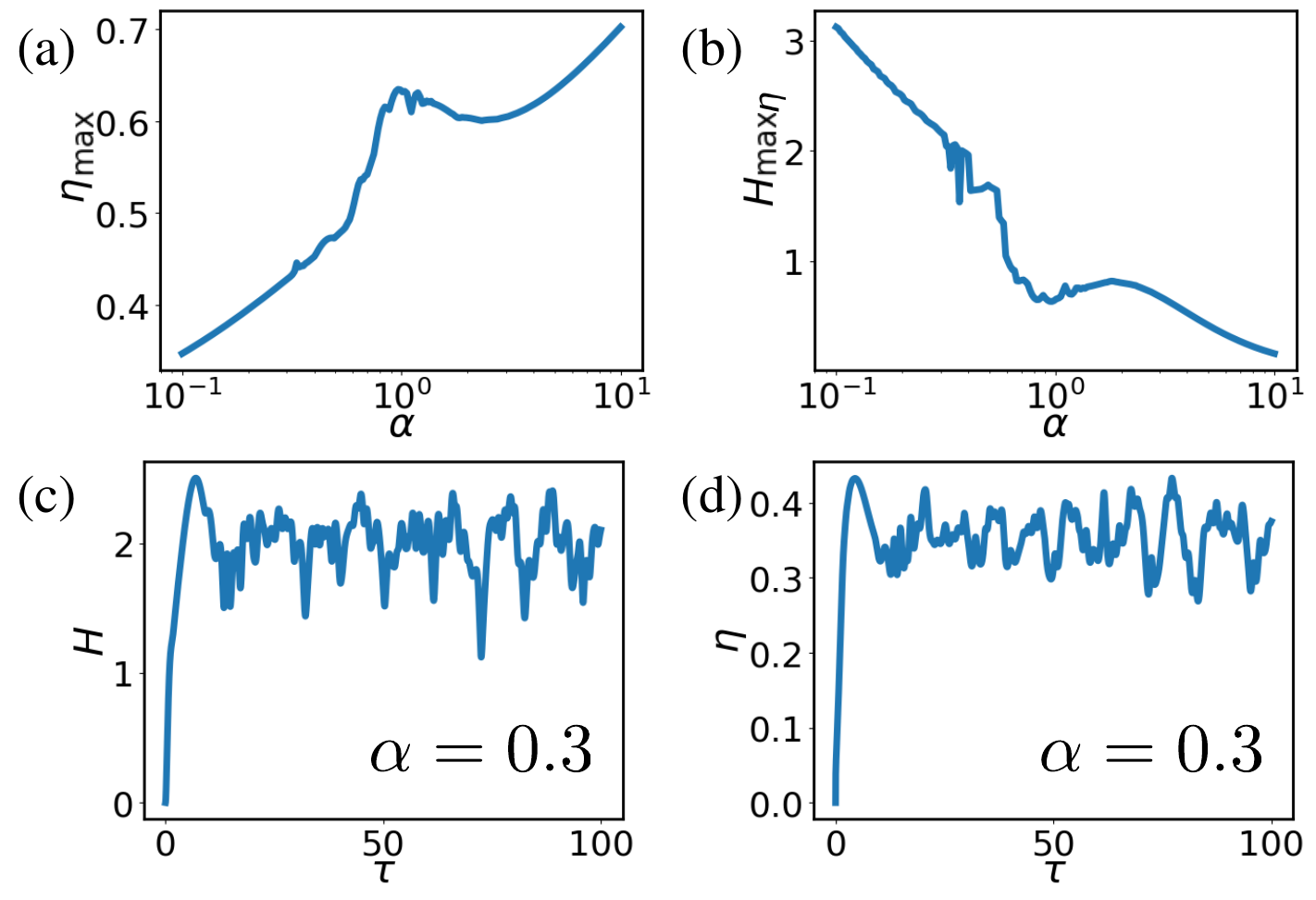}
    \caption{(a) Maximum efficiency for $\alpha\in[0.1, 10]$. We observe an increase as a bump with some irregular fluctuations. (b) Shannon entropy for the evolution time of $\tau^*_{\eta}(\alpha)$ showing an inverted irregular bump. (c) Dynamics of the Shannon entropy for $\alpha=0.3$ showing rapid and irregular fluctuations. (d) Highly fluctuating dynamics of the efficiency for $\alpha =0.3$.\label{fig: eta zoom in}}
\end{figure}

\subsection{An interpretation of the bump in the maximum efficiency}\label{app: bump}
We give a qualitative explanation for the appearance of the bumps in the maximum efficiency and the corresponding optimal time shown in Fig.~\ref{fig: efficiency}(b) and (c) in the main text. To see the bump clearly, we zoom in on an intermediate regime with $\alpha\in[0.1,10]$ for $\tilde{T}=1$. The maximum efficiency shown in Fig.~\ref{fig: eta zoom in}(a) increases around $\alpha =  O(1)$ with some irregular oscillations. In Fig.~\ref{fig: eta zoom in}(b), we plot the Shannon entropy $H$ at $\tau = \tau^*_{\eta}$ as a function of $\alpha$. A similar but inverted bump is seen in the same region, indicating that the wave function has a shrunk profile in this regime. According to the property of Bloch oscillations that an initially localized particle is constrained within a limited region, we consider the bump to be caused by a remnant of Bloch oscillations induced by a significant interference of waves due to $J$ and $\Delta$.

The oscillations on the bump can be attributed to the interference between Bloch oscillations and a wave reflected by the left rigid wall. What oscillates even more irregularly is the Shannon entropy $H$. We examine the dynamics of the Shannon entropy for $\alpha=0.3$ in this region. In Fig.~\ref{fig: eta zoom in}(c), we observe more irregular oscillations of $H$. To attain the maximum efficiency, we expect to have a small Shannon entropy. However, due to rapid fluctuations in the Shannon entropy, many local minima have similar values with very different evolution times. On the other hand, such fluctuations also greatly influence the numerical calculations because a slight deviation in the numerically found optimal time may result in a large deviation in the value of the Shannon entropy and energy gain, and consequently in the efficiency as shown in (d). This provides an interpretation of the irregular oscillations in the bumps. The underlying physics is understood as follows. In the small-$\alpha$ regime, the hopping by $J$ dominates the gradient $\Delta$. The wave packet travels in the right direction after a transient stage in which interference between the reflected wave by the left boundary occurs. The shape of the wave packet is maintained for a long time. In the large-$\alpha$ regime, hopping becomes weak and the dynamics is effectively described by rapid Rabi oscillations. For both cases, the dynamics of the wave packet is relatively regular. However, in the intermediate regime, the wave reflected by the boundary has a complicated interference with significant Bloch oscillations due to $J$ and $\Delta$, resulting in a highly fluctuating region.

\begin{figure}[t!]
    \centering
    \includegraphics[width=\columnwidth]{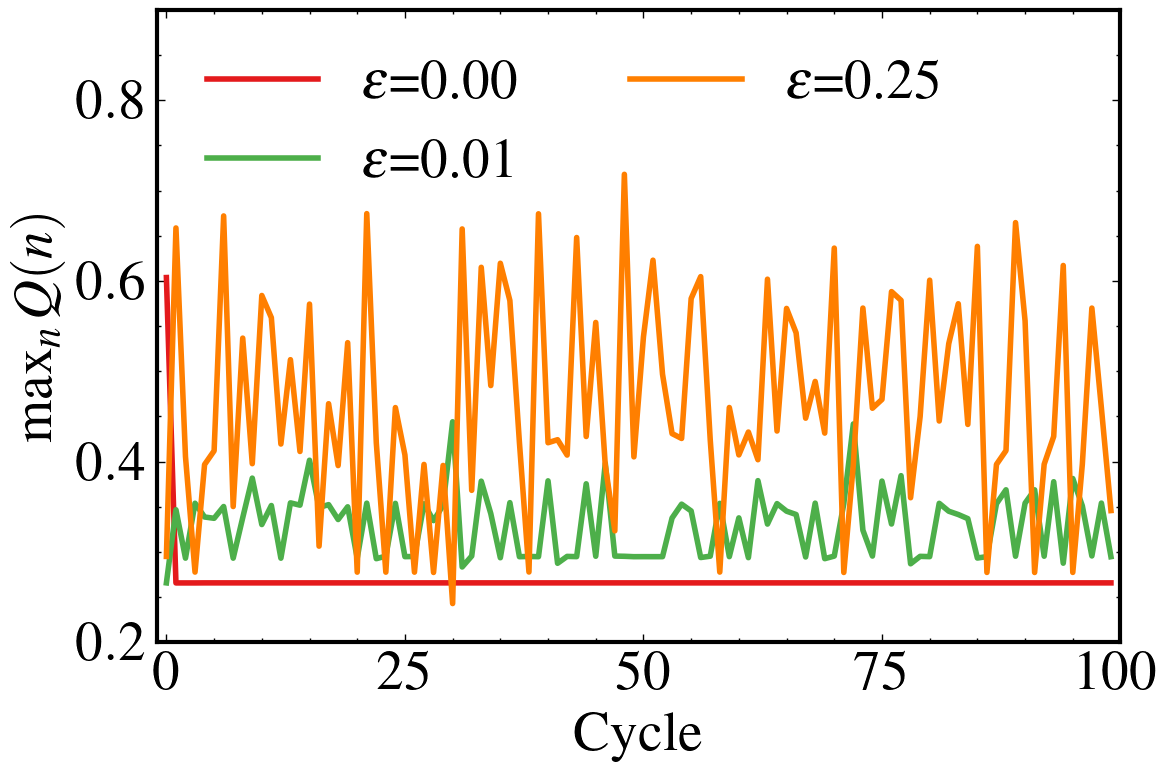}
    \caption{Numerical evidence of the disappearance of a steady state for the $j-2$ FB with a finite error $\varepsilon$ by showing the maximum population $\max_n Q(n,m\tau) := \max_n |\braket{n|\psi(m\tau)}|^2$ of the wave function of the time-evolved state $\ket{\psi(m\tau)}$ at each engine cycle labeled by an integer $0\le m \le M=100$ with $\alpha=1$. The red curve shows the error-free case with a constant value of $\max_n Q(n,m\tau)$, indicating a steady state. The green and orange curves plot the cases with $\varepsilon=0.01$ and $\varepsilon=0.25$, respectively with the non-converging $\max_n Q(n,m\tau)$, meaning a non-steady state.\label{fig: j-2_single_cycle_justfity}}
\end{figure}

\begin{figure*}[t!]
    \centering
    \includegraphics[width=\textwidth]{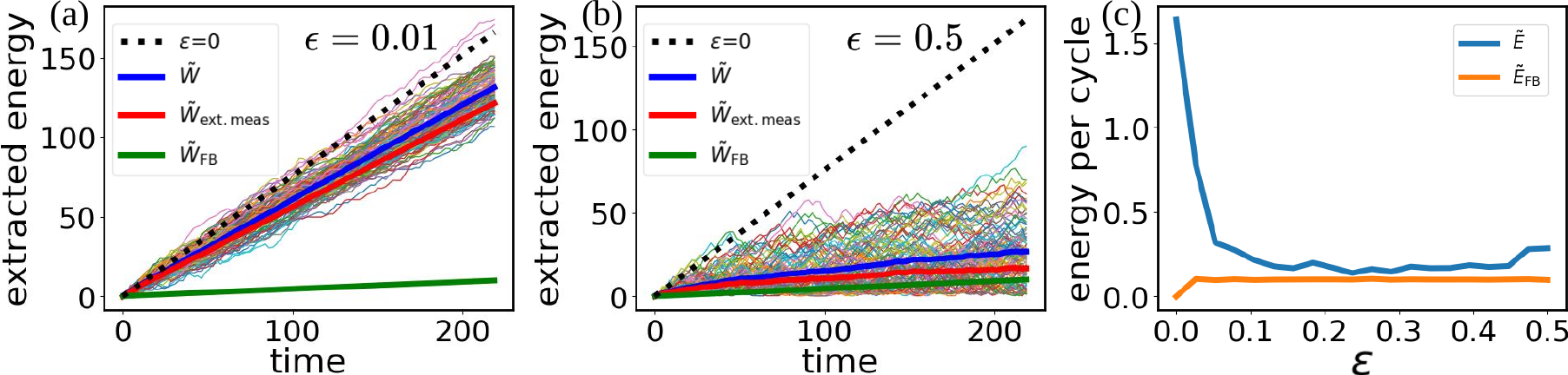}
    \caption{Total cumulative energy gain with a finite error $\varepsilon$ for the $j-1$ FB-controlled engine with $\alpha=1$. The evolution time for each cycle is the error-free optimal time $\tau=\tau^*_E (\alpha=1) = 2.21$. Each thin colored curve in (a) and (b) shows a trajectory of the total accumulated energy gain $\sum_{m=0}^M \tilde{w}_m$ over $M=100$ engine cycles with (a) $\varepsilon=0.01$ and (b) $\varepsilon=0.5$. The thick blue curves are the average total cumulative energy gain $\tilde{W}$ averaged over all trajectories. We use black dotted lines to represent the error-free results corresponding to $\tilde{E}_{\rm max}(1)=1.68 $. The average direct work injection $\tilde{W}_{\rm FB}$ performed by applying the FB is shown by the thick green curve. The thick red curve plots the average extracted work from quantum fluctuations through quantum measurements, given by $\tilde{W}_{\rm ext.meas} = \tilde{W} - \tilde{W}_{\rm FB}$. (c) Average total energy gain per cycle $\tilde{E}$ shown in blue and FB-injected work $\tilde{E}_{\rm FB}$ shown in orange as a function of $\varepsilon$.}
    \label{fig: weak_j-1_energy}
\end{figure*}

\section{Details for imprecise measurements}\label{app: j-2 FB}
\subsection{Breakdown of single-cycle calculations with $\varepsilon>0$}
In the precise measurement case, we optimize the performance within each FB cycle. However, numerical evidence shows that $\varepsilon>0$ makes single-cycle calculations insufficient. In Fig.~\ref{fig: j-2_single_cycle_justfity}, we plot for the $j-2$ FB the maximal population defined as $\max_n Q(n,m\tau) := \max_n |\braket{n|\psi(m\tau)}|^2$ of the time-evolved state $\ket{\psi(m\tau)}$ at each engine cycle labeled by an integer $0\le m \le M=100$. The time-evolved state at the $m$th cycle is given by
\begin{equation}
   \ket{\psi(m\tau)} := \exp(-i \tilde{H}_{j_m} \tau) \frac{M_{j_m}\ket{\psi((m-1)\tau}}{\sqrt{P(j_m)}},
\end{equation}
where $\ket{\psi(0)} :=\ket{g}$ is the ground state of the initial Hamiltonian $H_0$, $j_m$ is the measurement outcome obtained in the $m$th cycle, and $P(j_m)$ is the probability of this measurement outcome.

In the numerical simulations, we choose $\alpha=1$ and $\tau = \tau^*_p(\alpha=1)\approx 2.65$ for different values of the measurement error $\varepsilon$. The red curve in Fig.~\ref{fig: j-2_single_cycle_justfity} plots $\max_n Q(n,m\tau)$ versus $m$ for $\varepsilon=0$. Except for $m=1$, $\max_n Q(n,m\tau)$ quickly stabilizes to a constant value. This is because a precise measurement exactly prepares the same pre-evolving state for every engine cycle up to a shift in the site index. A stabilized maximum population is a necessary condition for the existence of a steady state up to a shift, thereby allowing a single-cycle calculation. However, with measurement errors, the post-measurement state generally depends on the state in the previous cycle because some quantum coherence persists. We introduce a small error $\varepsilon=0.01$ in the green curve and a larger error $\varepsilon=0.25$ in the orange curve. We see that the maximum population no longer stabilizes even for a small error. This means that an imprecise measurement \textit{no longer prepares the same pre-evolving state} for each cycle and a well-defined steady state does not exist. This fact hinders single-cycle analyses as done for precise measurements. It is necessary to run long trajectories and take the ensemble average over many trajectories. The breakdown of single-cycle calculations due to imprecise measurements also occurs for the $j-1$ FB for a similar reason.

\subsection{Relative contribution of work involved in the $j-1$ FB}\label{app:FB work}
To show the relative contribution of $\tilde{W}_{\rm FB}$ and $\tilde{W}_{\rm ext.meas}$, we numerically calculate $\tilde{W}$, $\tilde{W}_{\rm FB}$, and $\tilde{W}_{\rm ext.meas}$ for a small error $\varepsilon=0.01$ in Fig.~\ref{fig: weak_j-1_energy}(a) and a large error $\varepsilon=0.5$ in Fig.~\ref{fig: weak_j-1_energy}(b). Each thin colored curve represents a trajectory involving $M=100$ cycles. Here the evolution time for each cycle is chosen to be $\tau^*_E (\alpha=1) = 2.21$ which maximizes the average energy gain per cycle $\tilde{E}_{\rm max}(1)=1.68 $ in the error-free case depicted by the black dotted line. Taking the trajectory ensemble average yields $\tilde{W}$ as shown by the thick blue curve. The average FB-injected energy $\tilde{W}_{\rm FB}$ is plotted by the thick green line. 

For $\varepsilon=0.01$, the height of the wall $\tilde{V} = 10$ is not so small compared to the step height of $\alpha=1$. This ensures a not-so-large probability of tunneling to the left of the wall as can be seen in the trajectories in Fig.~\ref{fig: weak_j-1_energy}(a). The total energy gain $\tilde{W}$ includes a small contribution $\tilde{W}_{\rm FB}$ due to applying the FB and $\tilde{W}_{\rm ext.meas}$ as the main contribution. However, for $\varepsilon = 0.5$ as plotted in Fig.~\ref{fig: weak_j-1_energy}(b), the fraction of the contribution due to applying FB is large. The height of the wall $\tilde{V} = 0.5$ is smaller than the step height of $\alpha=1$. Therefore, the accumulated energy gain decreases significantly due to substantial tunneling through the wall potential to the left as observed in the trajectories in Fig.~\ref{fig: weak_j-1_energy}(b). Nevertheless, on average, our engine still gains energy with a low but nonzero power. In Fig.~\ref{fig: weak_j-1_energy}(c), we plot the numerical average of the total energy gain per cycle $\tilde{E}:=\tilde{W}/M$ shown in blue and the FB-injected work per cycle $\tilde{E}_{\rm FB}:=\tilde{W}_{\rm FB}/M$ shown in orange. It is clear that as long as $\varepsilon$ is within 5\%, the extracted energy from quantum fluctuations by Maxwell's demon is still a major contribution to the total energy gain.

\bibliography{ref}

\end{document}